\documentclass[reprint, amsmath, amssymb, aps, prx, nofootinbib]{revtex4-2}
\DeclareMathOperator*{\argmin}{\arg\!\min}
\usepackage{enumitem, framed, soul}
\usepackage{graphicx}
\usepackage{dcolumn}
\usepackage{bm}
\usepackage[backref=none]{hyperref} 
\usepackage[usenames,dvipsnames]{xcolor}
\usepackage[framemethod=TikZ]{mdframed}
\newcommand\myshade{85}
\colorlet{mylinkcolor}{violet}
\colorlet{mycitecolor}{YellowOrange}
\colorlet{myurlcolor}{Aquamarine}
\hypersetup{
  linkcolor  = mylinkcolor!\myshade!black,
  citecolor  = mycitecolor!\myshade!black,
  urlcolor   = myurlcolor!\myshade!black,
  colorlinks = true,
}

\begin{document}

\preprint{APS/123-QED}

\title[Wasserstein Solution Quality and the Quantum Approximate Optimization Algorithm: A Portfolio Optimization Case Study]{Wasserstein Solution Quality and the Quantum Approximate Optimization Algorithm: A Portfolio Optimization Case Study}

\author{Jack S. Baker$^{1}$}
\email{jack@agnostiq.ai}

\author{Santosh Kumar Radha$^{1}$}%
\email{santosh@agnostiq.ai}
\affiliation{$^1$Agnostiq Inc., 325 Front St W, Toronto, ON M5V 2Y1}

\date{\today}

\begin{abstract}
Optimizing of a portfolio of financial assets is a critical industrial problem which can be approximately solved using algorithms suitable for quantum processing units (QPUs). We benchmark the success of this approach using the Quantum Approximate Optimization Algorithm (QAOA); an algorithm targeting gate-model QPUs. Our focus is on the quality of solutions achieved as determined by the Normalized and Complementary Wasserstein Distance, $\eta$, which we present in a manner to expose the QAOA as a transporter of probability. Using $\eta$ as an application specific benchmark of performance, we measure it on selection of QPUs as a function of QAOA circuit depth $p$. At $n = 2$ (2 qubits) we find peak solution quality at $p=5$ for most systems and for $n = 3$ this peak is at $p=4$ on a trapped ion QPU. Increasing solution quality with $p$ is also observed using variants of the more general Quantum Alternating Operator Ans\"{a}tz at $p=2$ for $n = 2$ and $3$ which has not been previously reported. In identical measurements, $\eta$ is observed to be variable at a level exceeding the noise produced from the finite number of shots. This suggests that variability itself should be regarded as a QPU performance benchmark for given applications. While studying the ideal execution of QAOA, we find that $p=1$ solution quality degrades when the portfolio budget $B$ approaches $n/2$ and increases when $B \approx 1$ or $n-1$. This trend directly corresponds to the binomial coefficient $nCB$ and is connected with the recently reported phenomenon of \textit{reachability deficits}. Derivative-requiring and derivative-free classical optimizers are benchmarked on the basis of the achieved $\eta$ beyond $p=1$ to find that derivative-free optimizers are generally more effective for the given computational resources, problem sizes and circuit depths.
\end{abstract}

\maketitle

\section{Introduction \label{section:introduction}}

Mean-variance portfolio optimization (MVPO) is a problem at the foundations of Modern Portfolio Theory (MPT). Introduced in 1952 in the seminal work of Markowitz \cite{Markowitz1952}, MVPO is concerned with the selection of a portfolio of assets with maximal expected financial returns and minimal volatility given a predefined appetite for risk. Despite extensions to the more general post-MPT  \cite{Rom1993}, MVPO has retained its status as the workhorse for a breadth of users all the way from large financial institutions like commercial and investment banks down to small hedge funds and individual investors. In realistic formulations of MVPO, only discrete amounts of assets can be traded. In this situation, MVPO can be considered a quadratic unconstrained binary optimization (QUBO) problem which are in general NP-hard \cite{Pardalos1992}. \par

\begin{figure}
    \centering
    \includegraphics[width=\linewidth]{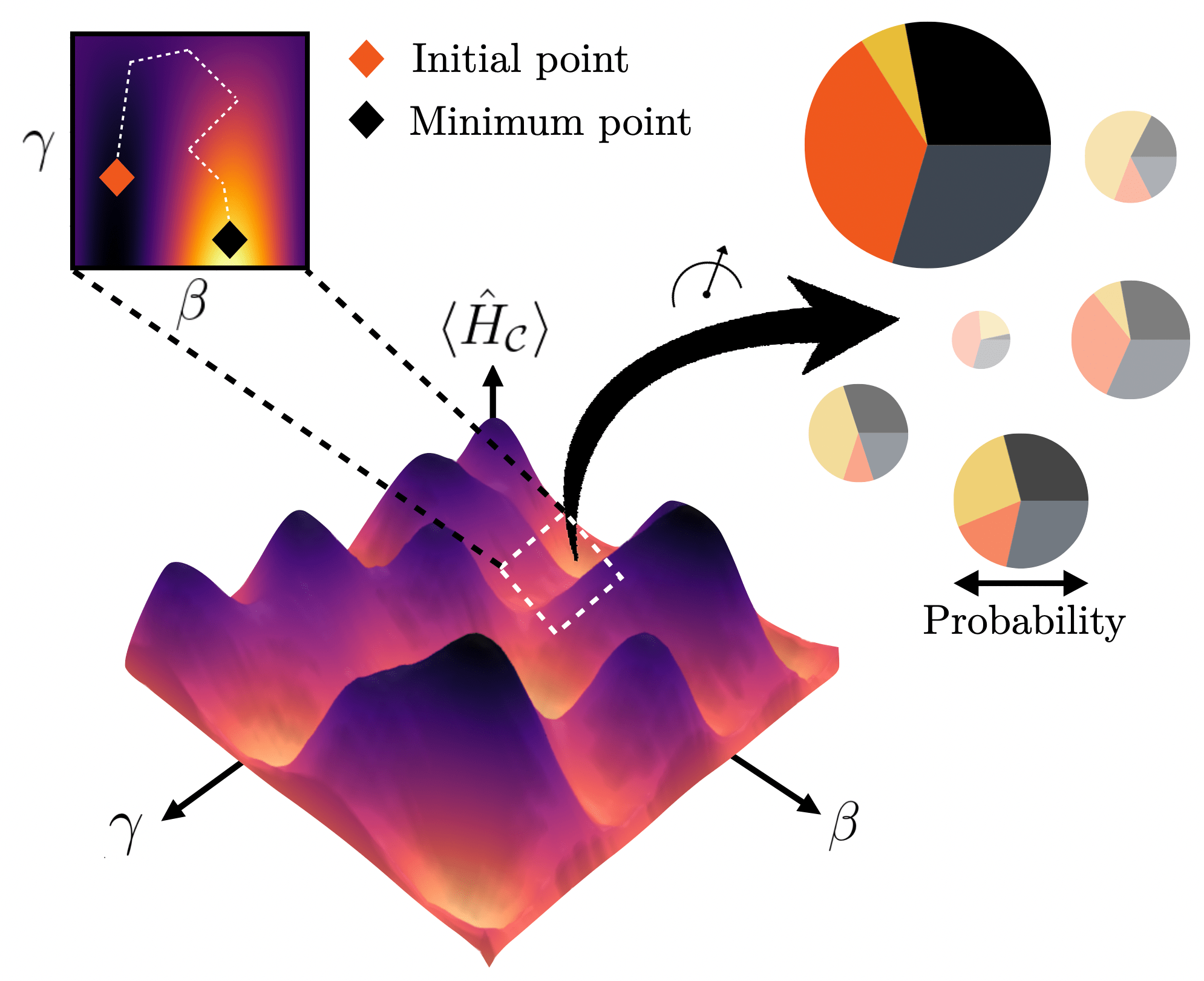}
    \caption{A schematic summarizing QAOA-based MVPO. An exemplar $\langle \hat{H}_{\mathcal{C}} \rangle (\gamma, \beta)$ (center) as evaluated using QPU calls at various parameters $\gamma$ and $\beta$. $\langle \hat{H}_{\mathcal{C}} \rangle (\gamma, \beta)$ is navigated by a classical optimizer along some path (white dotted line on inset) starting from some initial point to reach a minimum point. At this point, high quality portfolios are measured from the QPU with high probability.}
    \label{fig:schematic}
\end{figure}

While there now exists a plethora of classical heuristics and metaheuristics designed to solve QUBO problems approximately \cite{Glover1986, Glover1989, Glover1990, Beasley98heuristicalgorithms, Boros2007, Dunning2018, Aramon2019, Goto2021}, the pioneering works of Farhi introduced two (albeit related) quantum metaheuristics: Quantum Annealing (QA) \cite{Farhi2000} and the Quantum Approximate Optimization Algorithm (QAOA) \cite{Farhi2014}. Both algorithms are presently the topic of intense debate as to whether they will show performance improvements over classical approaches \cite{Hauke2020, Mandr2018, Guerreschi2019, farhi2019quantum, farhi2015quantum, barak2015beating}, possibly using near-future noisy intermediate scale quantum (NISQ) hardware. Indeed, QA is suitable for specialized quantum annealer machines which now scale to $\sim 1000$ (non-universal) qubits \cite{Bunyk2014, boothby2020nextgeneration, boothby2021architectural} when the QAOA is deployable to universal gate-model computers which, as of the time of writing, have not yet reached a comparable scale. As seeded by this scale discrepancy, QA-based MVPO has been the topic of several studies \cite{Marzec2013, Rosenberg2016, elsokkary2017financial, Venturelli2019, cohen2020portfolio40, cohen2020portfolio60, phillipson2020portfolio, Mugel2021, mugel2021dynamic} while QAOA-based MVPO \cite{hodson2019portfolio, Bartkoutsos2020, Bartschi2020, Egger2021, Slate2021} has received much less attention. So much so, that until the present work, systematic benchmarking of the QAOA-based approach has not been performed. \par

In this work, our focus is directed towards the QAOA (and its extension to the more general Quantum Alternating Operator Ans\"{a}tz \cite{Hadfield2019}) and its applicability for solving discrete MVPO. In order to build upon previous works \cite{hodson2019portfolio, Bartkoutsos2020, Bartschi2020, Egger2021, Slate2021}, one must consider thoroughly all of the working parts which comprise the hybrid quantum-classical nature of the QAOA. That is, the QAOA is a member of a class of variational algorithms \cite{McClean2016, Cerezo2021, bharti2021noisy} whereby the parameters of a quantum circuit are tuned by a classical optimizer to extremize an objective function (often the expectation value of some observable) evaluated using a quantum processing unit (QPU). Figure \ref{fig:schematic} illustrates this process with reference to MVPO. For the quantum aspect of this process, we must consider the effects of noise (shot-based or otherwise) and circuit depth as well as other facets of circuit design. Specifically, we must investigate different mixing Hamiltonians and state initializations which, together, define different constraint enforcement schemes vital to MVPO \cite{hodson2019portfolio}. On the classical side, there are a myriad of different selection strategies for the variational parameters of the quantum circuit. Some are general \cite{audet2017derivative, Cook2020, bergholm2020pennylane} while some are problem specific \cite{brandao2018fixed, Wurtz2021, wurtz2021classically, Wurtz2021P1}. This is a vital area to address as a recent work has suggested that this sub-task is NP-hard \cite{Bittel2021}. In this work, we focus on different flavours of ``black-box" optimization \cite{audet2017derivative}; optimizers which are not imbued with any special information about the objective function. We study well known optimizers in the derivative-free and derivative-requiring categories. We compare these strategies choosing to work within what we consider \textit{reasonable resources} given the present maturity/availability of gate-model hardware. For the resources used and problems studied, we find that derivative-free optimizers are on average the most effective. \par

Crucially, a complete set of application specific benchmarks should include the performance of different quantum backends. Broadly, these can be broken into two groups: quantum circuit simulators and real QPUs. Previous works on QAOA-based MVPO \cite{hodson2019portfolio, Bartkoutsos2020, Bartschi2020, Egger2021, Slate2021} have focused on the former and to the best of the author's knowledge, there have been no reported benchmarks for the latter. In this work, we study two varieties of quantum circuit simulator: dense statevector simulation and stochastic shot-based simulation \cite{viamontes2009quantum, Suzuki2021}. Both backends simulate the ideal execution of a gate model QPU but the former evaluates expectation values of observables with exact matrix algebra while the latter evaluates these expectation values with a finite number of measurements (shots) like a real QPU. These backends are useful to realize the ideal execution of different algorithmic variations with or without the presence of shot noise. For real QPUs, we use a selection of NISQ machines made available by IBMQ \cite{ibmq2021}, Rigetti \cite{rigetti2021} and IonQ \cite{ionq2021, Wright2019}. This forms a broad subset of available hardware spanning different hardware paradigms (superconducting and trapped ion qubits) and qubit connectivities. When measured against the most comparable (although none are directly comparable) of previous benchmarks \cite{otterbach2017unsupervised, qiang2018,alam2019analysis, Pagano2020, Willsch2020, Bengtsson2020, Abrams2020, Harrigan2021, weidenfeller2022scaling} we find increasing solution quality at problem sizes and circuit depths larger than have been previously reported as well as demonstrating the first deployment of the more general Quantum Alternating Operator Ans\"{a}tz to real QPUs. After finding a significant variability in solution quality between identical runs (above the level of shot-based noise) on all QPUs, we suggest that variability itself should be provided as a benchmark for all QPUs specific to individual applications.

Given all of the above considerations, we must carefully select a metric to form a basis for our benchmarks. While time to solution is critical, in this work, our focus is on an equally important factor: the quality of solutions achievable given identical classical and quantum computational effort. To measure the quality of solutions, we reformulate the Wasserstein distance (WD) approach introduced in \cite{radha2021quantum}. We describe how this metric is able to provide a means of comparing the success of different QAOA runs, independent of the numerical details of a given problem instance and of the constraint enforcement scheme chosen. This is in contrast to the popular approximation ratio approach which is sensitive to all of these factors. We present the WD metric in a way which exposes the role of the QAOA as a transporter of probability and show empirically for which problem instances this transport becomes \textit{harder}. Notably, under the ideal execution of the QAOA, we find that the WD metric degrades when the portfolio budget $B \approx n/2$, increasing when $B$ approaches $1$ or $n-1$. This trend matches the evolution of the binomial coefficient $nCB$ with $B$ and is related to the recently reported phenomenon of \textit{reachability deficits} \cite{Akshay2020, zhang2021quantum}. In totality, this work can be considered a collection of \textit{application specific benchmarks} for QAOA-based MVPO on real QPUs and ideal quantum circuit simulators; an area which is attracting an increasing amount of attention in several different areas in quantum computation \cite{lubinski2021applicationoriented, Proctor2021, Amoretti2021, Mills2021}. 

The rest of this work is now organized as follows. In Section \ref{sec:qaoa_qubo_ising}, we present the MVPO formalism and show how its discrete variation takes the form of a QUBO/Ising objective. We also detail the basics of the QAOA, its extensions and different methods for enforcing the budget constraint in MVPO. In Section \ref{sec:NCWD}, we present in a pedagogical way the WD metric for solution quality first proposed in \cite{radha2021quantum} and discuss the advantages of using it and its variations. In Section \ref{sec:ideal_simulator}, we study the ideal execution of QAOA-based MVPO on quantum circuit simulators first on random problem instances at $p=1$ (Section \ref{subsec:singledepth}) and then on representative problem instances obtained using real market data at $1 \leq p \leq 5$ (Section \ref{subsec:classicalopt}), detailing the performance of different classical optimizers. In Section \ref{sec:realqpu}, we use the optimal ans\"{a}tze (the sns\"{a}tze using the optimal parameters) obtained in Section \ref{subsec:classicalopt} to prime a limited re-optimization of the variational parameters on a selection of real gate-model QPUs. We study each QPU with a focus on the quality of solutions achieved and their fidelity to ideal execution (Section \ref{subsec:perform_vs_depth_real}) as well as studying the variability in solution quality between identical runs (Section \ref{subsec:variability}). We finish in Section \ref{sec:conclusions} summarizing this work and providing some insights into the future research directions for QAOA-based MVPO. 

\section{MVPO and the QAOA \label{sec:qaoa_qubo_ising}}

\subsection{Canonical MVPO \label{subsec:mvpo}}

MVPO is concerned with the Markowitz objective $\mathcal{C}(\boldsymbol{w})$ 

\begin{equation}
\mathcal{C}(\boldsymbol{w}) = (1 - \lambda)\boldsymbol{w} \boldsymbol{\Sigma} \boldsymbol{w}^T - \lambda \boldsymbol{\mu} \cdot \boldsymbol{w} 
\label{eq:continuous_markowitz}
\end{equation}

and its minimization with respect to the portfolio weights vector $\boldsymbol{w}$ of an $n$-asset pool while constrained to a budget $B$

\begin{equation}
\min \left\{\mathcal{C}(\boldsymbol{w}) \mathrel{\bigg|} \sum_{i=0}^{n-1} w_i = B \right\}
\label{eq:min_markowitz}
\end{equation}

to find the optimal portfolio weights vector $\boldsymbol{w}^*$

\begin{equation}
\boldsymbol{w}^* = \argmin [\mathcal{C}(\boldsymbol{w})]. 
\label{eq:argmin_markowitz}
\end{equation}

In Equation \ref{eq:continuous_markowitz}, $0 \leq \lambda \leq 1$ is the risk appetite factor, $\boldsymbol{\Sigma}$ is the $\mathbb{R}^{n \times n}$ covariance matrix and $\boldsymbol{\mu}$ is the $\mathbb{R}^{n}$ vector of expected returns. In a simple sense, MVPO is the act of finding a balance between portfolio volatility (the first term of Equation \ref{eq:continuous_markowitz}) and expected portfolio returns (the second term of Equation \ref{eq:continuous_markowitz}). Given historical market data (thus $\boldsymbol{\mu}$ and $\Sigma$), the precise point of balance is determined by $\lambda$. Should $\boldsymbol{w}^*$ be achieved, this portfolio is said to sit on the efficient frontier \cite{Markowitz1952, Rom1993}

While $w_i$ and B need not take on specific bounds, in the canonical case, $-1 \leq w_i \leq 1$ and $B = 1$. This formulation is a continuous optimization problem, which, as previously noted, does not account for the realistic scenario where only discrete amounts of assets can be traded. The next Section presents a simple (and pedagogical) discrete formulation of MVPO.

\subsection{Discrete MVPO as a QUBO/Ising problem \label{subsec:mvpo_as_qubo}}

\begin{figure}
    \centering
    \includegraphics[width=\linewidth]{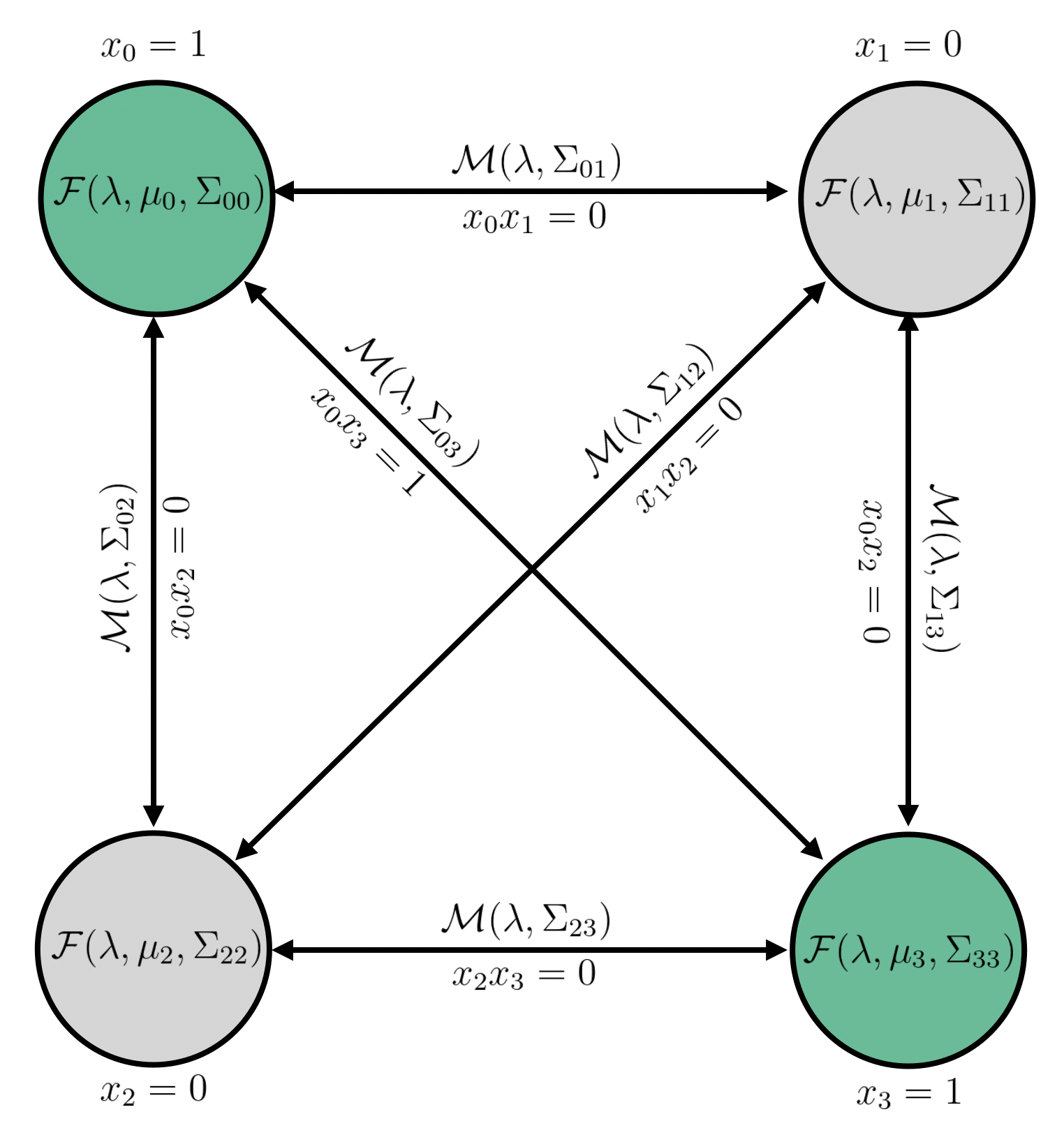}
    \caption{$n=4$ Buy/hold MVPO represented on the complete graph $K_4$ with bit vector $\boldsymbol{x} = [1, 0, 0, 1]$. Selected vertices (assets, $x_i=1$) are colored in green while those not selected ($x_i = 0$) are colored in grey. Linear terms $\mathcal{F}(\lambda, \mu_i, \Sigma_{ii}) = (1 - \lambda)\Sigma_{ii} - \lambda \mu_i$ and quadratic terms $\mathcal{M}(\lambda, \Sigma_{ij}) = (1 - \lambda)\Sigma_{ij}$ are shown over their appropriate vertices and edges.}
    \label{fig:graph_mvpo}
\end{figure}

Reference \cite{hodson2019portfolio} used the QAOA to solve discrete MVPO with ternary portfolio weights: $w_i \in \{-1, 0, 1\}$, or, sell, hold or buy. This formulation requires $2n$ qubits and uses a bit string encoding scheme which permits degenerate solutions. We choose to simplify this approach by using only binary portfolio weights $w_i \in \{0, 1\}$ and $B \in \mathbb{Z}_0^+$, $0 \leq B \leq n$. This defines a buy/hold (1/0) trading strategy with a strictly non-degenerate objective function provided the elements of $\boldsymbol{\mu}$ and the independent elements of $\boldsymbol{\Sigma}$ are unique. For notational convenience, for buy/hold MVPO we relabel the portfolio weights $w_i \rightarrow x_i \in \{0, 1\}$. This way, it is clear that $\mathcal{C}(\boldsymbol{x})$ takes on the QUBO form

\begin{equation}
    \mathcal{C}(\boldsymbol{x}) = \sum_{i=0}^{n-1} \sum_{j=0}^i Q_{ij} x_i x_j
    \label{eq:qubo_objective}
\end{equation}

for QUBO matrix elements $Q_{ij}$. In this form, buy/hold MVPO can be represented on the complete graph $K_n$ as is shown in Figure \ref{fig:graph_mvpo}. Assets are represented as vertices and selecting one ($x_i = 1$) incurs a linear term. Selecting any pair of vertices ($x_ix_j = 1$) incurs a quadratic term. Now, given that we must select $B$ vertices, we must find the combination of vertices where the value of the incurred terms is minimal. The QUBO form of Equation \ref{eq:qubo_objective} is important since $C(\boldsymbol{x})$ is always re-writable in terms of spin variables $s_i \in \{-1, 1\}$ and the corresponding spin vector $\boldsymbol{s}$. That is, upon substituting $x_i = (s_i + 1)/2$ we obtain a quadratic Ising model

\begin{equation}
    \mathcal{C}(\boldsymbol{s}) = \sum_{i=0}^{n-1} \sum_{j=i+1}^{n-1} J_{ij} s_i s_j + \sum_{i=0}^{n-1} h_i s_i + c
    \label{eq:ising_objective}
\end{equation}

for quadratic coupling matrix elements $J_{ij}$, linear coefficients $h_i$ and constant term $c$. The constant term need not be considered in any optimization problem since it introduces no relative shift between the cost of any solution. It is of course required to recover the value of $\mathcal{C}(\boldsymbol{s})$. In this work, the transform between binary and spin variables is performed using \texttt{PyQUBO} (\texttt{v1.0.5}) \cite{zaman2021pyqubo}. As we shall describe in Section \ref{subsec:qaoa_intro}, our transform to spin variables allows us to encode $\mathcal{C}(\boldsymbol{s})$ into a Hamiltonian which is diagonal in the computational basis. \par

\subsection{The Canonical QAOA \label{subsec:qaoa_intro}}

The QAOA is a metaheuristic algorithm for approximately solving combinatorial optimization problems using gate-model quantum computers. This Section provides a brief summary of the original or \textit{canonical} form of the algorithm first proposed by Farhi \cite{Farhi2014}.

First, we turn our attention back to Equation \ref{eq:ising_objective}; the quadratic Ising model. Its form implies that $\mathcal{C}(\boldsymbol{s})$ is the sum of $m$ local terms or \textit{clauses} where no term involves more than two spins

\begin{equation}
    \mathcal{C}(\boldsymbol{s}) = \sum_{\alpha=1}^{m}\mathcal{C}^{\alpha}(\boldsymbol{s}).
    \label{eq:n_clauses}
\end{equation}

Further, because each spin variable $s_i$ is an eigenvalue of the Pauli-$z$ matrix $\hat{\sigma}^z_i$, $\mathcal{C}(\boldsymbol{s})$ [and $\mathcal{C}^{\alpha}(\boldsymbol{s})$] can be recast into an operator form, $\hat{H}_{\mathcal{C}}$ (and $\hat{H}^{\alpha}_{\mathcal{C}}$), which is diagonal in the computational basis. Using this form, we define a parameterized unitary

\begin{equation}
    \hat{U}_{\gamma} = e^{-i \gamma \hat{H}_{\mathcal{C}}} = \prod_{\alpha = 1}^m e^{-i \gamma \hat{H}_{\mathcal{C}}^{\alpha}}
    \label{eq:cost_unitary}
\end{equation}

where the product form is possible since each $\hat{H}_{\mathcal{C}}^{\alpha}$ commutes with each other. Equation \ref{eq:cost_unitary} is often known the cost unitary or phase separation unitary. \par

Now, we define a simple mixing operator (an $x$-mixer)

\begin{equation}
    \hat{U}_{\beta} = e^{-i \beta \sum_{j=0}^{n-1} \hat{\sigma}_j^x} = \prod_{j=0}^{n-1} e^{-i \beta \hat{\sigma}_{j}^{x}},
    \label{eq:xmixer}
\end{equation}

possible once again since Pauli-$x$ matrices commute with each another. We note that this form of $\hat{U}_{\beta}$ mixes all states and is only one choice from a family of mixers which can be engineered to interact with only a certain subspace of states. Other choices for this unitary will be discussed in Section \ref{subsec:qaoa_extension}. \par

Before applying the above unitary transformations, states are initialized as the uniform superposition

\begin{equation}
    |+ \rangle ^{\otimes n} = \frac{1}{\sqrt{2^n}} \sum_{x} |x\rangle
    \label{eq:hadamardstate}
\end{equation}

where the sum is over all permutations of the $n$-bit basis state $| x \rangle$ (here, $x$ is the bit string representation $x_0 x_1 \cdots x_{n-1}$ of the bit vector $\boldsymbol{x}$). The algorithm now proceeds by applying a $p$-layered unitary operator to $|+ \rangle^{\otimes n}$ to obtain a parameterized quantum state $|\boldsymbol{\gamma}, \boldsymbol{\beta} \rangle$

\begin{equation}
    |\boldsymbol{\gamma}, \boldsymbol{\beta} \rangle = \left[ \prod_{k=1}^{p}\hat{U}_{\gamma_p}\hat{U}_{\beta_p} \right]|+\rangle ^{\otimes n}
    \label{eq:gamma_beta_state}
\end{equation}

where $\boldsymbol{\gamma}$ and $\boldsymbol{\beta}$ are parameter vectors of length $p$: $\boldsymbol{\gamma} = [\gamma_1, \gamma_2, \cdots, \gamma_p]$, $\boldsymbol{\beta} = [\beta_1, \beta_2, \cdots, \beta_p]$. In this state, we take the expectation of the observable $\hat{H}_{\mathcal{C}}$

\begin{equation}
    \langle \hat{H}_{\mathcal{C}} \rangle = \langle \boldsymbol{\gamma}, \boldsymbol{\beta} | \hat{H}_{\mathcal{C}} | \boldsymbol{\gamma}, \boldsymbol{\beta} \rangle.
    \label{eq:expectedcost}
\end{equation}

Equation \ref{eq:expectedcost} now becomes the variational objective of a classical minimization routine; $\boldsymbol{\gamma}$ and $\boldsymbol{\beta}$ are varied to minimize $\langle \hat{H}_C \rangle$. Should the global minimum of this objective function be found, we yield the minimum expected cost $M_p$ and the \textit{best} parameter vectors $\boldsymbol{\gamma}^*$ and $\boldsymbol{\beta}^*$ which achieved it. The quantum adiabatic theorem \cite{Farhi2000, Farhi2014} is able to give limits on the success of this minimization. That is

\begin{equation}
    \lim_{p \to \infty} M_p = \min{[\mathcal{C}(\boldsymbol{x})]} = \mathcal{C}(\boldsymbol{x}^*).
    \label{eq:adiabatic_theorem}
\end{equation}

 So, not only can the minimum of a classical objective be achieved at the limit of $p$, when increasing $p$, $M_p$ edges closer towards $\mathcal{C}(\boldsymbol{x}^*)$. This is because $M_{p-1}$ can be viewed as a constrained minimization of $M_p$ where the two new variational parameters are fixed at zero: $M_{p-1} \geq M_{p}$.

\subsection{Extensions to the QAOA and enforcing constraints \label{subsec:qaoa_extension}}

Since the original proposal of the QAOA \cite{Farhi2014}, the algorithm was extended to the more general Quantum Alternating Operator Ans\"atz \cite{Hadfield2019} (from hereon also referred to as the QAOA). Above all, this approach allows more general unitaries to be used than those presented in Section \ref{subsec:qaoa_intro}, which, when paired with a suitable state initialization, allow for alternate constraint enforcement schemes. Before detailing this, however, let us consider how constraints can be enforced without changing the formalism in Section \ref{subsec:qaoa_intro}. \par

As seen in Equation \ref{eq:min_markowitz}, MVPO requires the enforcement of a single equality constraint. To do so, we can add a penalty term $\mathcal{K}$ to $\mathcal{C}(\boldsymbol{x})$

\begin{equation}
    \mathcal{K} = \alpha \left( \sum_{i=0}^{n-1} x_i - B \right)^2
    \label{eq:penalty_term}
\end{equation}

where $\alpha$ is a penalty scaling factor. This approach is called \textit{soft constraints} because the full space of solutions are still considered; constraint violating solutions are merely forced to be disfavored. It is clear that $\alpha$ must be at least large enough to yield the correct hierarchy of solutions to $\mathcal{C}(\boldsymbol{x})$ (solutions which satisfy constraints are favoured to those which do not). Setting $\alpha$ to the minimal value to achieve this was suggested in \cite{hodson2019portfolio}. In this work, we test two approaches for setting $\alpha$: the minimal value method suggested in \cite{hodson2019portfolio} ($\alpha = \alpha_{min}$) and setting $\alpha$ arbitrarily high ($\alpha = 100$).

Now, we demonstrate how one can proceed without using a penalty term under the new formalism: \textit{hard constraints}. We begin with a state initialization which includes only those states which satisfy the equality constraint. The simplest choice is a random state with Hamming weight $B$. This is the state $|R_n^B \rangle$ where $R_n^B$ is any randomized bit string of length $n$ whose sum of all bits (the Hamming weight) is equal to $B$. This is implemented simply by applying $\hat{\sigma^x}$ to $B$ random qubits. Another way is prepare a Dicke state

\begin{equation}
    | D^n_B \rangle = \frac{1}{\sqrt{\binom{n}{B}}} \sum_{|\boldsymbol{x}| = B} | x \rangle.
    \label{eq:dicke_state}
\end{equation}

This is an equally weighted superposition of all states which satisfy the equality constraint. In this work, we use the efficient preparation scheme proposed in \cite{Brtschi2019} to prepare Dicke states with $\mathcal{O}(Bn)$ gates and $\mathcal{O}(n)$ circuit depth without the use of ancilla qubits. \par

Now with a constraint satisfying initial state, we must use a mixing Hamiltonian which (unlike Equation \ref{eq:xmixer}) only mixes states within the viable subspace. One such choice of Hamiltonian for achieving this is a ring mixer

\begin{equation}
    \hat{H}_R = \sum_{i=0}^n \hat{\sigma}^x_i \hat{\sigma}^x_{i+1} + \hat{\sigma}^y_i \hat{\sigma}^y_{i+1}
    \label{eq:ring_mixer_hamiltonian}
\end{equation}

for Pauli-$x$ and Pauli-$y$ matrices $\hat{\sigma}^x_i$ and $\hat{\sigma}^y_i$ acting on the $i^{\text{th}}$ qubit, where qubits are arranged in periodic boundary conditions (i.e, a ring; the index $i + 1$ is taken $\mod n$). Equation \ref{eq:ring_mixer_hamiltonian} can be viewed as a SWAP operation between neighbouring pairs of qubits. This operation conserves the Hamming weight of any bit string so our search remains only in the viable subspace of states. Another choice is the complete graph mixer whose construction is similar to  Equation \ref{eq:ring_mixer_hamiltonian} but the sum is over all pairs of qubits on the complete graph $K_n$

\begin{equation}
    \hat{H}_K = \sum_{(i, j) \in E(K_n)} \hat{\sigma}^x_i \hat{\sigma}^x_{j} + \hat{\sigma}^y_i \hat{\sigma}^y_{j}.
    \label{eq:complte_graph_hamiltonian}
\end{equation}

Previous works have shown the superiority of the complete graph mixer over the ring mixer under the ideal execution of QAOA for graph coloring \cite{Wang2020} and ternary portfolio optimization \cite{hodson2019portfolio}. Although for specific cases (see the max-$\kappa$-colorable-subgraph problem, for example \cite{Wang2020}) the ring and complete graph mixers have been realized exactly with circuits logarithmic and linear in $n$, respectively, presently, a general implementation of both mixers requires a truncated Trotter expansion. To keep the depth of such an implementation low, we choose a low order expansion with a fixed Trotter step size $\epsilon = 0.25$. While we accept that this choice may introduce noticeable Trotter error, when implemented on real NISQ hardware, using a a deeper circuit (more terms in the Trotter expansion) will likely incur signifficant coherent and incoherent noise which will dominate over the reduced Trotter error. 

In this work, we use two different implementations of hard constraints. The first uses a Dicke state initialization alongside the complete graph mixer ($|D_n^B \rangle$ \& $\hat{H}_{K}$) and the second uses random Hamming states alongside the ring mixer ($|R_n^B \rangle$ \& $\hat{H}_{R}$). From previous results on other optimization problems \cite{hodson2019portfolio, Wang2020}, we expect the former approach to yield the best results under the ideal execution of the QAOA, but, since the second requires fewer total gates, it is unclear which approach will perform best on presently available NISQ hardware.

\section{Measuring solution quality 
\label{sec:NCWD}}

In this Section, we present pedagogically an alternate way of thinking about the quality of solutions achieved by any combinatorial optimization algorithm whose result is a probability distribution of solutions. Our method is based upon the concept of Wasserstein distances and is valid for constrained and unconstrained problems. We introduce it now in a way which is most relevant to the QAOA. \par

After the completion of a QAOA run, provided the classical optimization step converged to the global minimum of $ \langle H_{\mathcal{C}} \rangle$, we obtain $M_p$. This can be written in terms of $\mathcal{C}(\boldsymbol{x})$ and a discrete probability distribution $P^*(\boldsymbol{x})$; the probability of measuring the bit vector $\boldsymbol{x}$ from a measurement of the optimal ans\"atz

\begin{equation}
    M_p = \sum_{l=0}^{2^n-1} \mathcal{C}(\boldsymbol{x}_l) P^*(\boldsymbol{x}_l)
    \label{eq:optimal_cost}
\end{equation}

where the sum is in principle over every possible bit vector, but, in real QPU usuage (or shot-based simulation), is only over those bit vectors which were measured in the finite number of shots. In the context of the above, the categorical distribution $P^*(\boldsymbol{x})$ can be used interchangeably with the ordinal distribution $P^*[\mathcal{C}(x)]$ since we do not require a notion of distance between bins. From here on-wards, we drop the super script $^*$ for probability distributions $P$ for notational clarity. Now, it is commonplace in the literature to use $M_p$ to measure the success of a QAOA run through calculating the approximation ratio $r$

\begin{equation}
    r = \frac{M_p}{\mathcal{C}(\boldsymbol{x}^*)}
    \label{eq:approx_ratio}
\end{equation}

or the 0-1 bounded variation $r^b$

\begin{equation}
    r^b = \frac{M_p - \mathcal{C}(\boldsymbol{x}^{\text{max}})}{\mathcal{C}(\boldsymbol{x}^*) - \mathcal{C}(\boldsymbol{x}^{\text{max}})}, \quad 0 \leq r^b \leq 1
    \label{eq:approx_ratio_bounded}
\end{equation}

where $\boldsymbol{x}^{\text{max}}$ is the bit vector which maximizes $\mathcal{C}(\boldsymbol{x})$. While $r$ and $r^b$ are useful measures, they suffer similar flaws. It is apparent that $r$ and $r^b$ must depend on the precise definition of $\mathcal{C}$. Importantly, if we are using QAOA to optimize a constrained problem, the form of $\mathcal{C}$ will vary depending on the method used to enforce constraints (i.e, with or without the penalty term of Equation \ref{eq:penalty_term}). Using soft constraints, $\mathcal{C}$ will feature an additional dependency on $\alpha$. It is clear that the choice of $\alpha$ influences $\mathcal{C}$ and therefore $r$ (or $r^b$). The problem with this observation is clear should we consider two penalty factors $\alpha_1$ and $\alpha_2$, $\alpha_1 \neq \alpha_2 \neq 0$ in an otherwise identical $\mathcal{C}$. After performing a QAOA run using each $\alpha$, let's say we achieved $P_1(\boldsymbol{x}) = P_2(\boldsymbol{x})$. Despite yielding the same probability distribution, Equation \ref{eq:optimal_cost}, \ref{eq:approx_ratio} and \ref{eq:penalty_term} show that a different $r$ is achieved; if a solution exists satisfying the Hamming weight constraint, the denominator of Equation \ref{eq:approx_ratio} is unchanged (with no $\alpha$ dependence) while the numerator is an increasing function of $\alpha$. We clearly cannot fairly compare the solution quality of runs 1 and 2 using $r$. Moreover, if we instead used hard constraints, there is no need for the term $\mathcal{K}$ (thus $\alpha=0$) and we once again cannot make fair comparisons between hard and soft schemes using $r$. Similar arguments can be used to show that $r^b$ also depends (differently) on $\alpha$ and thus cannot be used for fair comparison between constraint enforcement schemes. \par

Therefore, to make fair comparisons between different constraint enforcement schemes, we must construct a different solution quality metric agnostic to the precise numerical details of $\mathcal{C}$. We begin first by splitting the $2^n$ bit vectors $\boldsymbol{x}$ into two sets: those which satisfy the constraints (viable), and those which do not (not viable). In the buy/hold MVPO case, sets are grouped by the equality constraint on $B$: $X^v = \{\boldsymbol{x}: \sum_{i=0}^{n-1} x_i = B \}$ and $X^{nv} = \{\boldsymbol{x}: \sum_{i=0}^{n-1} x_i \neq B \}$ respectively. From these two sets, we construct the two ascending order sequences $\mathcal{C}^{v} = (\mathcal{C}(\boldsymbol{x}): \boldsymbol{x} \in X^v)$ and $\mathcal{C}^{nv} = (\mathcal{C}(\boldsymbol{x}): \boldsymbol{x} \in X^{nv})$\footnote{Another route is to create a larger number of sets grouped by the magnitude of constraint violation: $X^{nv}_1 = \{\boldsymbol{x}: \sum_{i=0}^{n-1} x_i = B \pm 1 \}$, $X^{nv}_2 = \{\boldsymbol{x}: \sum_{i=0}^{n-1} x_i = B \pm 2 \} \ldots$ which can be used to build the corresponding ascending order sequences $\mathcal{C}^{nv}_1$, $\mathcal{C}^{nv}_2 \ldots$ which as before are concatenated to the end of $\mathcal{C}^{v}$. Indeed, in the case of soft constraint MVPO, this is equivalent to building a single sequence $\mathcal{C}^{all} = (\mathcal{C}(\boldsymbol{x}): \boldsymbol{x} \in X^{v} \cup X^{nv})$.}. At this point, we make it clear that if we were to study a different and unconstrained problem, we can continue with $\mathcal{C}^{nv} = ()$, the empty sequence. Should our problem have constraints (like MVPO), following the discussion in Section \ref{subsec:qaoa_extension}, one may wonder why $\mathcal{C}^{nv}$ need be considered at all for hard constraints since we operate only within the viable subspace. However, this is only true for the ideal execution of the algorithm and not for execution on NISQ hardware where probability can ``leak" in to the unviable space of solutions. We now proceed by concatenating the two sequences to form $\mathcal{G} = \mathcal{C}^{v} {^\frown} \mathcal{C}^{nv}$. Within this sequence, we find the index corresponding to the cost $\mathcal{C}(\boldsymbol{x})$. This is given by the inverse

\begin{equation}
    \tilde{\mathcal{C}}(\boldsymbol{x}) \equiv \mathcal{G}^{-1}(\mathcal{C}(\boldsymbol{x})), \quad 0 \leq \tilde{C} \leq 2^n -1, \quad \tilde{C} \in \mathbb{Z}^+_0.
    \label{eq:rank}
\end{equation}

Within this formalism, $\tilde{\mathcal{C}}(\boldsymbol{x}^*) = 0$. In simple terms, what we have created is a function which when given $\boldsymbol{x}$ returns the ranking of $\boldsymbol{x}$ as judged by (i) whether $\boldsymbol{x}$ is a viable solution to the constrained optimization problem and (ii) the value $\mathcal{C}(\boldsymbol{x})$. The next step is to use $\tilde{\mathcal{C}}$ as bin labels of a new probability distribution $P(\tilde{\mathcal{C}})$. In doing so, we have transformed from ordinal to interval variables in $P$; the distance between bins in now unity and does not depend on the numerical details of $\mathcal{C}$. \par

\begin{figure}
    \centering
    \includegraphics[width=\linewidth]{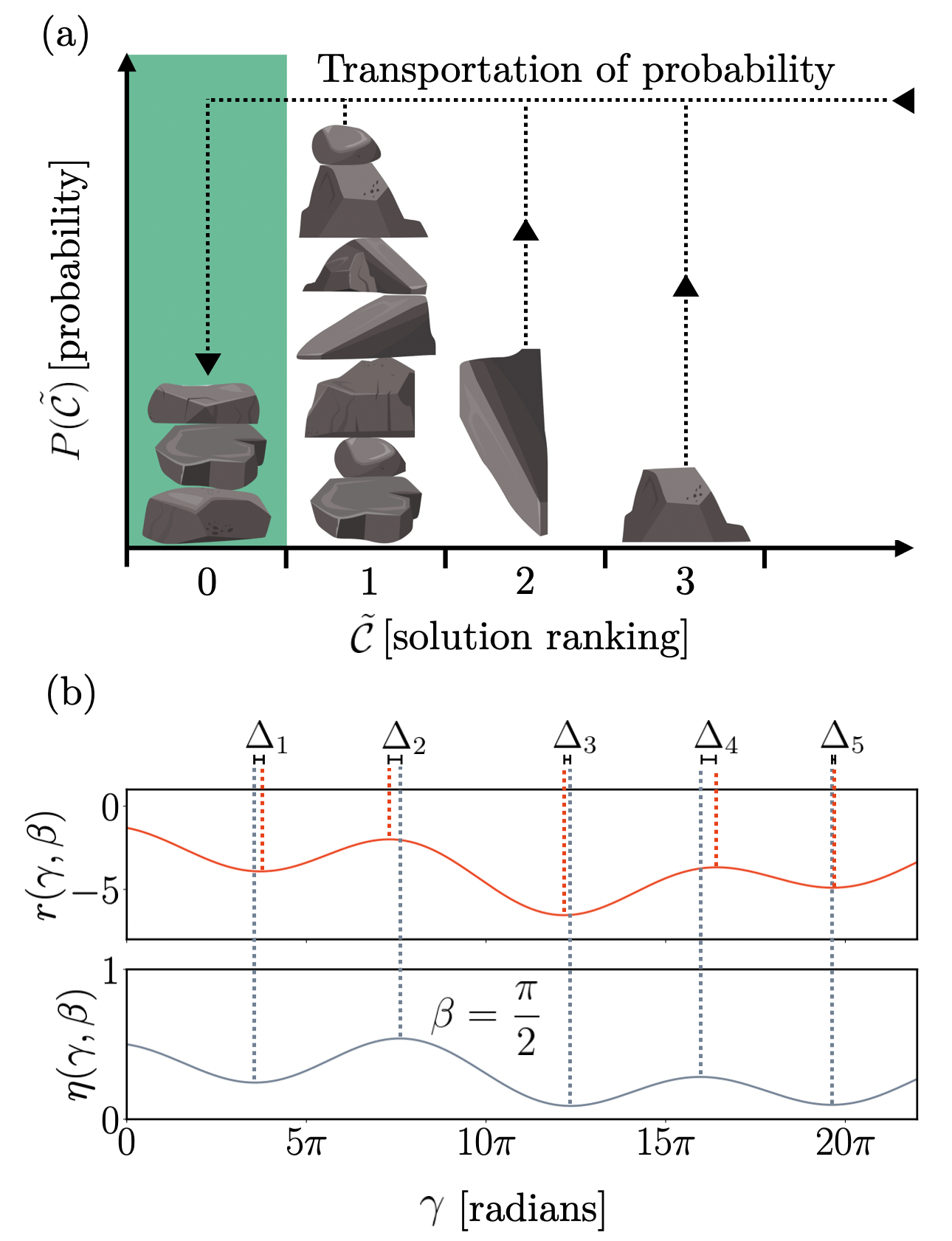}
    \caption{(a) A schematic describing the Normalized and Complementary Wasserstein Distance, $\eta$, in direct analogy with the earth-mover nature of the metric. All of the solutions to $\mathcal{C}(\boldsymbol{x})$ are given a ranking $\tilde{\mathcal{C}}$ (see Equation \ref{eq:rank}) which are used as bins for the probability distribution $P(\tilde{\mathcal{C}})$. A given $P(\tilde{\mathcal{C}})$ has a large value of $\eta$ if little work must be done to transport all probability (shown as rocks, in line with the earth-mover analogy) from $\tilde{\mathcal{C}} \neq 0$ to $\tilde{\mathcal{C}} = 0$. In this case, we say that the solutions are of high quality. Other cases are discussed in the text. Individual rock images are taken from a royalty free repository \cite{rockcartoon}. (b) Empirically, $r$ and $\eta$ share the same quantity of minima, but their precise positions in $\gamma$ (and in $\beta$, but not shown here) differ by some amount $\Delta_i$  as shown above (b).}
    \label{fig:NCWDexplained}
\end{figure}

Given this ranking, we now ask a question using a simple analogy to classical physics: how much work must be done to transport all of the probabilities of measuring non-optimal $\tilde{\mathcal{C}}(x) \neq 0$ to the optimal $\tilde{\mathcal{C}}(\boldsymbol{x}^*) = 0$? This analogy is depicted in Figure \hyperref[fig:NCWDexplained]{3(a)} where rocks are transported to the origin. Should these rocks move against a gravitational field (along the $x$-axis of Figure \hyperref[fig:NCWDexplained]{3(a)}, say) more work must be done to transport them to the origin if they are further away. Replacing the weight of rocks for probability and the distance travelled for the solution ranking, we have

\begin{equation}
    \mathcal{W} = \sum_{l=0}^{2^{n}-1} \tilde{\mathcal{C}}(\boldsymbol{x}_l) P[\tilde{C}(\boldsymbol{x}_l)].
    \label{eq:work_done}
\end{equation}

Equation \ref{eq:work_done} is known as a Wasserstein-1 distance $W_1$ or \textit{earth mover} distance. Formally,

 \begin{equation}
     \mathcal{W} :=  W_1(\delta_{\tilde{\mathcal{C}}(\boldsymbol{x}), \tilde{\mathcal{C}}(\boldsymbol{x}^*)}, P[\tilde{\mathcal{C}}(\boldsymbol{x}])
 \end{equation}
 
 for Kronecker delta function $\delta_{\tilde{\mathcal{C}}(\boldsymbol{x}), \tilde{\mathcal{C}}(\boldsymbol{x}^*)} = 1, \text{if } \boldsymbol{x} = \boldsymbol{x}^*$, 0 otherwise. This is the form given in \cite{radha2021quantum}.

One choice is to normalize $\mathcal{W}$ by a factor $1/(2^{n} - 1)$ and take its complement. This leaves us with the normalized and complementary WD (NCWD) $\eta$

\begin{equation}
    \eta = 1 - \frac{\mathcal{W}}{2^n - 1}, \quad 0 \leq \eta \leq 1, \quad \eta \in \mathbb{R}_{\geq 0}.
    \label{eq:eta}
\end{equation}

Whether to use $\eta$ or $\mathcal{W}$ is completely context dependent. Indeed, in this work we make most use of $\eta$, but if one wants to study how solution quality scales with problem size (see Section \ref{subsubsec:soln_qual_constraints}), one may find $\mathcal{W}$ more instructive.

Proceeding with the NCWD, we note three important limits: (i) when $P(0) = 1$, the metric is maximal at $\eta = 1$, (ii) when $P(2^n-1) = 1$, the metric is minimal at $\eta = 0$ and (iii) when $P(0) = P(1) = P(2), ... = P(2^n-1) = 1/2^n$, $\eta = 0.5$, the results are indistinguishable from random. We also must emphasize that while $r$ and $\eta$ do bare some similarities, they are separate measures. One important difference is that the two measures do not share optima in their landscapes over $\boldsymbol{\gamma}$ and $\boldsymbol{\beta}$. This point is illustrated in Figure \hyperref[fig:NCWDexplained]{3(b)} where both measures are plotted as a function of $\gamma$ at fixed $\beta = \pi/2$ for a $p=1$ problem. It is empirically observed that there are small differences in optima positions. This means that for any given $M_p$ and optimal angles $\boldsymbol{\gamma}^*$ and $\boldsymbol{\beta}^*$, we can only guarantee that a global minimum of $r$ has been found, and not the global minimum of $\eta$. We finally remark upon the computational requirements of calculating $\eta$. Similarly to the calculation of $r$/$r^b$, we must perform an exhaustive search to find $\mathcal{C}(\boldsymbol{x}^*)$ [and for $r_b$, also $\mathcal{C}(\boldsymbol{x}^{\text{max}})$]. Considering a soft constraint objective, this scales $\mathcal{O}(2^n)$ in time. For $\eta$, however, we are also required to store the $2^n$ solutions in memory. Using $\eta$ to benchmark the success of the QAOA on real QPUs or quantum circuit simulators does therefore require that the problem size be small enough such that an exhaustive search can be used. Importantly, this means that once QPUs are capable of treating problem sizes beyond classical computation, $\eta$, $r$ and $r^b$ can no longer be used and a new benchmarking approach must be developed.

\section{Solution quality using ideal quantum circuit simulators \label{sec:ideal_simulator}}

\subsection{Random problem instances at $p = 1$ \label{subsec:singledepth}}

We begin by surveying the achievable solution quality for QAOA-based buy/hold MVPO at the minimal circuit depth $p=1$. We do so in consideration of a selection of random problem instances in $2 \leq n \leq 10$ scanning over $1 \leq B \leq n - 1$ for each $n$. Within these bounds, we generate 200 random instances of: $\boldsymbol{\Sigma}$ matrices, $\boldsymbol{\mu}$ vectors and values of $\lambda$. A large value of $\alpha = 100$ is used to enforce soft constraints. For the classical optimization step, we perform a search on a uniformly fine $50 \times 25$ grid of $\gamma_1$ and $\beta_1$ in the range $[0, 2\pi) \times [0, \pi)$. This choice of classical optimizer limits the effect of performance variation between different problem instances; it is plausible that other types of optimizer could perform better on the cost landscapes defined by some problem instances compared to others. After optimization, the $\boldsymbol{\gamma}$ and $\boldsymbol{\beta}$ which best minimize  $\langle \hat{H}_{\mathcal{C}} \rangle$ are used to calculate $P(\boldsymbol{x})$. For each random problem instance, we also perform a $2^{n}$ scaling brute force search of all solutions to $\mathcal{C}(\boldsymbol{x})$ which (following the discussion in Section \ref{sec:NCWD}) allows for the calculation of $\eta$ (Equation \ref{eq:eta}). Since we wish to estimate the ideal performance at $p = 1$, these simulations use the dense (exact) statevector simulator variation of \texttt{Qulacs} \cite{Suzuki2021} [\texttt{v0.3.0}]. \par

\begin{figure}
    \centering
    \includegraphics[width=\linewidth]{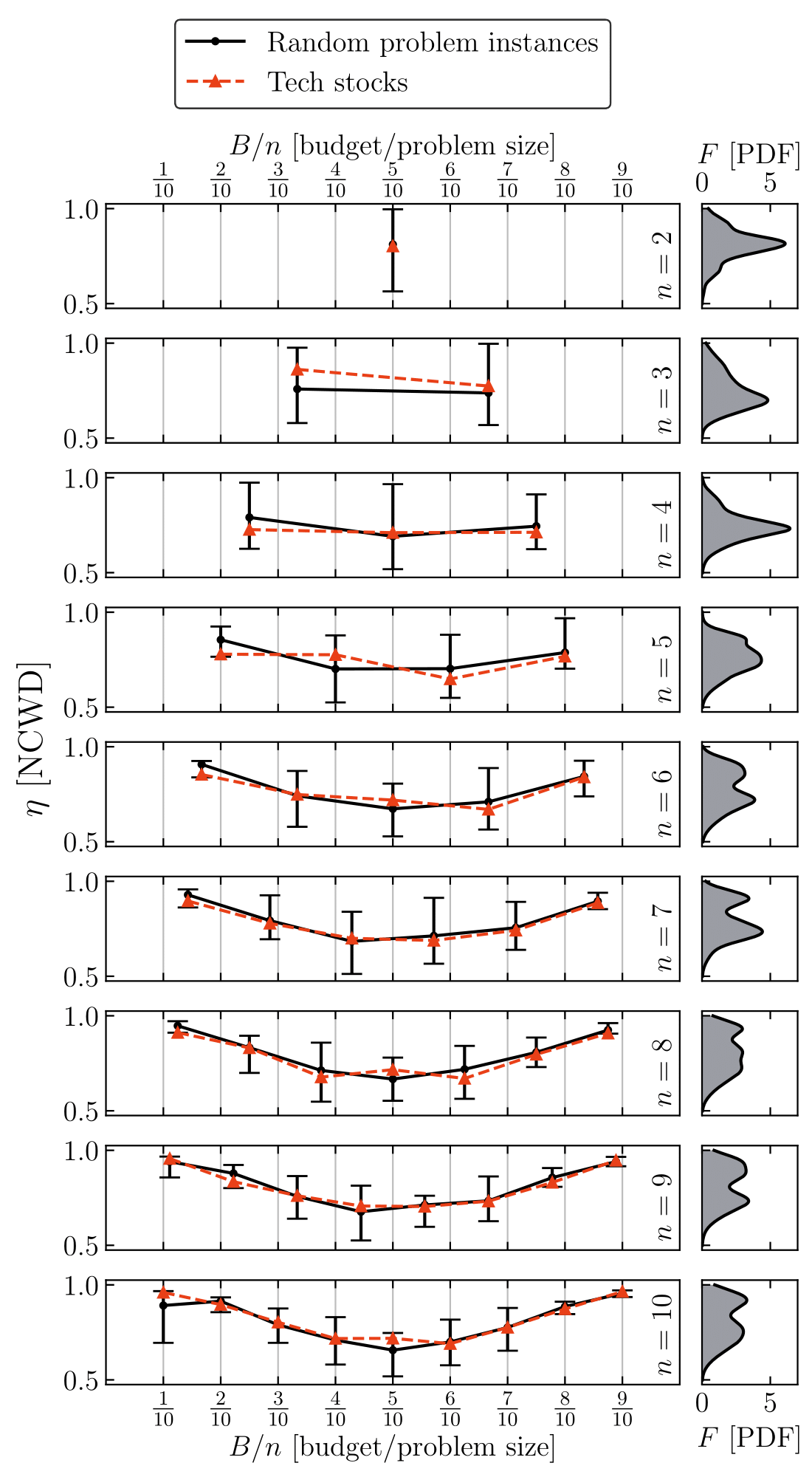}
    \caption{$\eta (B/n)$, $2 \leq n \leq 10$, $1 \leq B \leq n - 1$ using soft constraints with $\alpha = 100$. Random problem instances are shown with black solid line and circle marker. Error bars indicate the total spread. Selected tech stocks (see Section \ref{subsec:classicalopt}) are shown with a red dotted line and triangle marker. The PDF of $\eta$ over the 200 random problem instances, $F$, is shown for each $n$.}
    \label{fig:BbyN_pdf}
\end{figure}

The results are shown in Figure \ref{fig:BbyN_pdf}. For each $B/n$, we can see that there is a large (and variable) spread in $\eta$. Further analysis reveals that the spread is approximately normally distributed. Beyond $n=5$, we can also observe a tendency for extremal values of $B/n$ (i.e, 1 or $n-1$) to be more tightly grouped around the mean. The most important observation of these simulations is the trend in the mean of $\eta$ as a function of $B/n$. We can see that extremal $B$, on average, yields high quality solutions more often ($\eta$ closer to 1) while those $B/n$ closer to 0.5 yield a distribution of solutions closer to random more often ($\eta$ closer to 0.5). At $n \geq 4$, this leads to the development of distinct peaks in the probability density function $F$ (PDF; Figure \ref{fig:BbyN_pdf}, right) as driven by the different $B/n$. At the level of smearing provided, most PDFs feature two peaks. The low $\eta$ peak is created by intermediate $B/n$ while the upper $\eta$ peak is from extremal $B/n$.

\begin{figure}
    \centering
    \includegraphics[width=\linewidth]{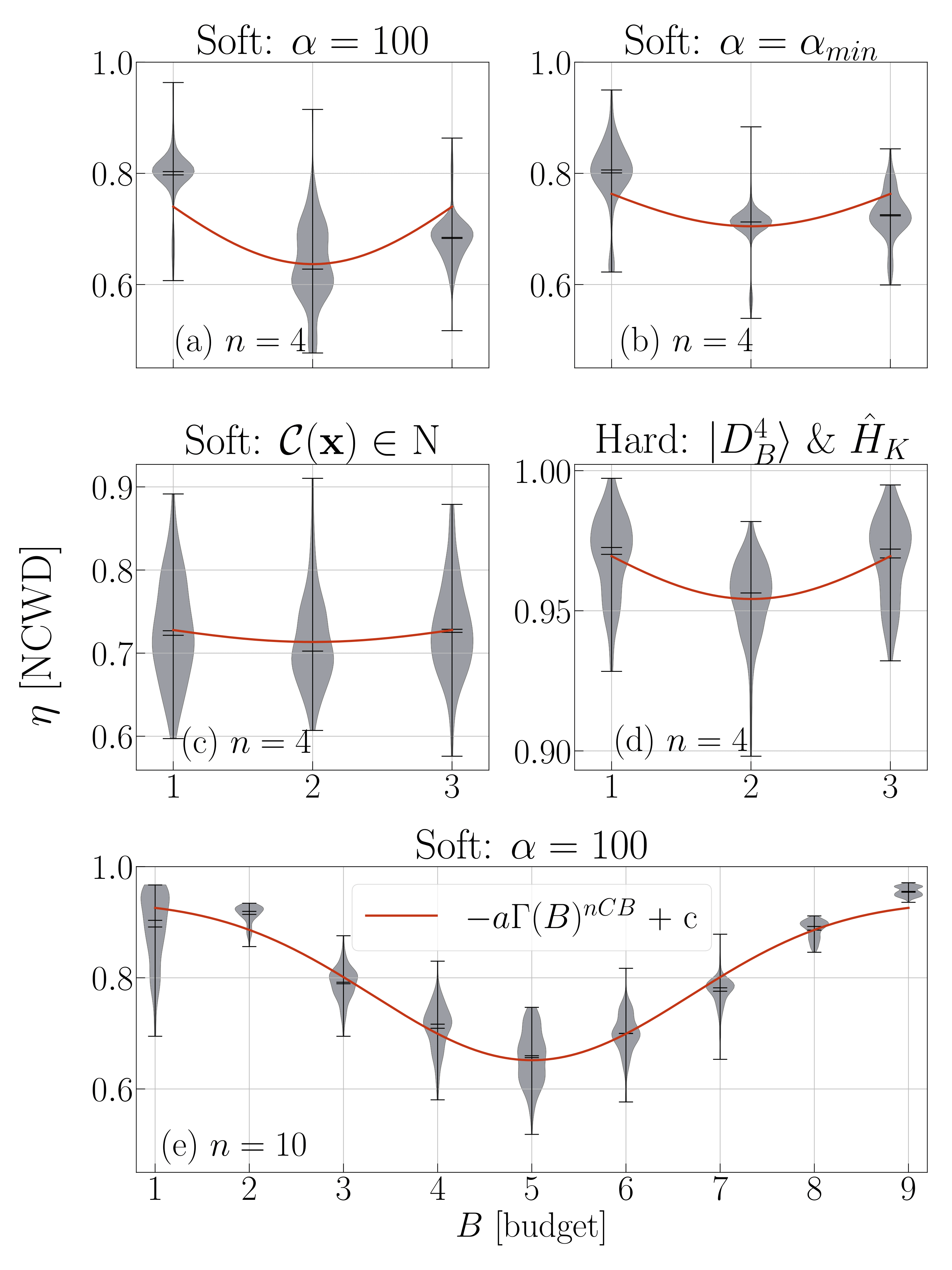}
    \caption{$\eta (B)$ for $1 \leq B \leq n - 1$. Each panel features 200 random problem instances for each $B$ and the trend has been fit using $-a\Gamma(B)^{nCB} + c$ where $a$ and $b$ are constants and $\Gamma(B)^{nCB}$ is defined in Equation \ref{eq:gamm_func}. Panels (a), (b) and (d) share a $y$-scale while (c) and (d) do not. This is to increase the visibility of the $\eta(B)$ trend in the mean. (a) Soft constraints using $\alpha = 100$ at $n=4$. (b) Soft constraints using $\alpha = \alpha_{min}$ at $n=4$. (c) Soft constraints using $\alpha=10,000$ restricting $\mathcal{C}(\boldsymbol{x}) \in \mathbb{N}$ at $n=4$. (d) Hard constraints as implemented using Dicke state initialization and the complete graph mixer at $n=4$. (e) Soft constraints using $\alpha=100$ for the $n=10$ problem.}
    \label{fig:gammafit}
\end{figure}

To consolidate the existence of observed trend in $\eta (B/n)$, we perform three more experiments. Focusing only on $n = 4$, we use 200 random problems instances to calculate $\eta(B)$ for (i) soft constraints with $\alpha = \alpha_{\text{min}}$, (ii) soft constraints where we limit $\mathcal{C}(\boldsymbol{x}) \in \mathbb{N}$, setting $\alpha =$ 10,000 and (iii) hard constraints using Dicke state initialization (Equation \ref{eq:dicke_state}) and the complete graph mixer (Equation \ref{eq:complte_graph_hamiltonian}); $|D_B^4 \rangle$ \& $\hat{H}_K$. The results are shown in Figure \ref{fig:gammafit}. Let us now motivate each experiment and detail their results. \par 

The experiment with $\alpha = \alpha_{\text{min}}$ is performed to demonstrate that the trend in $\eta(B/n)$ persists independently of the choice of $\alpha$ within the soft constraint formalism. Figure \hyperref[fig:gammafit]{5(a)} shows the original results for $\alpha=100$ and \hyperref[fig:gammafit]{5(b)} shows the results for $\alpha=\alpha_{\text{min}}$. It is clear that the mean $\eta$ follows a comparable trend in the two cases. The only difference is the mean $\eta$ is marginally higher for $\alpha=\alpha_{\text{min}}$ and $\eta$ is more widely spread for $\alpha = 100$. The experiment with $\mathcal{C}(\boldsymbol{x}) \in \mathbb{N}$ is necessary to rule out the observed trend being an artefact of $M_p$ not being enclosed within the range of the grid search. While the range of the grid in $\beta_1$ is sufficient for this, we cannot guarantee that $M_p$ be (approximately) found for a general problem since the bounds of $\gamma_1$ are determined by the eigenvalue spectrum of $\hat{H}_{\mathcal{C}}$ (or, equivalently, the QUBO/Ising parameters \cite{ozaeta2021expectation}). $\mathcal{C}(\boldsymbol{x}) \in \mathbb{N}$ is a special case where the eigenvalue spectrum leads to $\gamma_1$ becoming bound by $[0, 2\pi)$. This experiment therefore guarantees that $M_p$ is enclosed within the space of the grid search. Figure \hyperref[fig:gammafit]{5(c)} shows the results of this experiment; clearly the general observed trend persists, but is dampened. This dampening is from the more even distribution of $\eta$ in the spread of each $B/n$ point. We attribute this to the use of larger $\alpha$, which increases the roughness of the cost landscape to become like those observed in \cite{ozaeta2021expectation}, requiring the use of a finer grid search to find good minima. The last experiment using $|D_B^4 \rangle$ \& $\hat{H}_K$ shows that the observed trend persists in hard constraints. In Figure \hyperref[fig:gammafit]{5(d)}, it can be seen that we achieve a much more tightly grouped $\eta$ (note the scale $\eta$) for all $B/n$ with an enhanced solution quality compared to soft constraints. \par

Now we have verified the robustness of the $\eta(B/n)$ trend, we hypothesize its correspondence with a function in elementary combinatorics: the binomial coefficient $nCB$. We write this function in a form continuous in $B$ (with constant $n$) using Bernoulli $\Gamma$-functions 

\begin{equation}
    \Gamma^{nCB}(B) = \frac{\Gamma (n+1)}{\Gamma (B+1) \Gamma(n - B + 1)}
    \label{eq:gamm_func}
\end{equation}

where $\Gamma^{nCB}(B)= nCB$ when $n \in \mathbb{Z}^+$ and $B \in \mathbb{Z}^+$. Taking a function of the form $-a\Gamma^{nCB}(B) + c$ where $a$ and $b$ are constants, we fit the mean $\eta(B)$ for each $n$. A subset of the obtained functions are shown overlaid on Figure \ref{fig:gammafit} (red lines). For all cases, we see that the general trend in $\eta(B)$ is well reproduced by the fitted function. This trend bares some resembelence the phenomenon of \textit{reachability deficits} that has been reported for the QAOA applied to boolean satisfiability problem \cite{Akshay2020, zhang2021quantum}. That is, for the 3-SAT and 2-SAT problems, the quality of the approximate solutions returned by the QAOA was identified as an decreasing function of the problem density (equivalently, as given in \cite{Akshay2020}, the reachability deficit $\langle \hat{H}_{\mathcal{C}} \rangle - \mathcal{C}(\boldsymbol{x}^*)$ is an increasing function of problem density); the ratio of the number of clauses $m$ (see Equation \ref{eq:n_clauses}) to the problem size $n$: $m/n$. \par

Indeed, it is not directly possible to connect ours results with \cite{Akshay2020} and \cite{zhang2021quantum} since in our case the number of clauses is always fixed, regardless of the choice of $B$ or constraint enforcement scheme. In hard constraints, however, since our choice of state initialization and mixing Hamiltonian means we operate only within the space of viable solutions, some clauses are not interacted with by the algorithm. In that sense, our minimum of solution quality observed near $B/n = 0.5$ is at the peak of the problem density. In other words, the peak value of the binomial coefficient $nCB$ (the number of viable solutions) corresponds to a minimum of solution quality as given by the QAOA. Our observation of the same trend for soft constraints is not explainable in the same way. In this case, the number of clauses are constant and all clauses are interacted with. Changing $B$ in this case merely affects the number of eigenvalues of $\hat{H}_{\mathcal{C}}$ which are penalized, which, as our results show still gives rise to reachability deficits. Indeed, a much earlier paper \cite{hogg1996quantum} observed that the success of another quantum combinatorial search algorithm evolved with $m/n$ thus showing this phenomenon is widespread even beyond the QAOA.

\subsection{Real market data beyond $p=1$ \label{subsec:classicalopt}}

Having now investigated the performance of $p=1$ QAOA on a large number of random problems, we move to study a select number of problems beyond $p=1$ as generated using real market data. We study a single problem instance at each problem size in $2 \leq n \leq 10$ with QAOA circuit depths of $1 \leq p \leq 5$. As motivated by the discussion in Section \ref{subsec:singledepth}, we choose the $B$ which (on average) yields the lowest $\eta$. For problems with even $n$, this is $B = n/2$. For odd $n$ values of $3$, $5$, $7$ and $9$ we choose $B$ as $2$, $3$, $3$ and $4$, respectively. This choice provides insights into the lower bounds of the performance of the QAOA. We choose $\lambda$ to be 0.5 in all problems but we view this choice as arbitrary as we did not observe a strong dependence of $\lambda$ in Section \ref{subsec:singledepth}. We calculate the returns vectors and covariance matrices using historical market data from 2021/04/01 to 2021/04/30 (YYYY/MM/DD) inclusive for a selected number of stock tickers in the sequence $\mathcal{T} =$ (GOOG, AMZN, FB, NVDA, TSLA, AAPL, PYPL, MSFT, BABA, INTC) where $\mathcal{T}$ is listed descending order of returns over the specified time window. We select $n$ tickers from $\mathcal{T}$, choosing to remove tickers from the end of $\mathcal{T}$ when $n < 10$. \par

To check these selected problems can be deemed representative of the random problems studied in Section \ref{subsec:singledepth}, we first proceed by evaluating the solution quality achieved from $p=1$ problem using the gird search classical optimization method from Section \ref{subsec:singledepth}. The results of this step are overlaid on Figure \ref{fig:BbyN_pdf} with red dashed lines and triangle markers. While we observe some deviation from the mean solution quality of the random problems, the selected problems do fall within the spread of the random instances 100\% of the time. We therefore infer that our selected problem instances are not outliers and can in that sense be considered representative problems. We do note, however, that this definition of representability accounts only for the magnitude of optima (i.e, how deep are the minima of $\eta$?) and not for other characters of the cost landscape (like roughness) which could influence the performance of the classical optimizers treated in the proceeding Sections. These considerations are beyond the scope of this work and should be the topic of a separate and future work. 

\subsubsection{Black box classical optimization \label{subsubsec:blackbox}}

In this Section, we provide a brief summary of each of the classical optimizers we use in our benchmarks. They are provided below along with their original references if applicable: \par \medskip

\begin{mdframed}[roundcorner=10pt]

{\setlength{\parindent}{0cm}
\underline{\textit{Random Search}} - All $\gamma_p$ and $\beta_p$ are selected randomly from $[0, 2\pi )$. \par
}

\medskip

{\setlength{\parindent}{0cm}
\underline{\textit{Nelder-Mead}} \cite{Nelder1965} - A derivative-free downhill simplex algorithm. For the $2p$-dimensional optimization problem, a simplex of $2p + 1$ test points is constructed. Depending on the flavor of Nelder-Mead being used, the behaviour of $\langle \hat{H}_{\mathcal{C}} \rangle$ is measured using each point on the simplex which then by some flavor-dependent logic updates the points of the simplex. We use the version implemented in \texttt{SciPy} (\texttt{v1.5.2}) \cite{2020SciPy-NMeth}. \par
}

\medskip

{\setlength{\parindent}{0cm}\ul{\textit{Constrained Optimization By Linear Approximations (COBYLA)}} \cite{Powell1994} - A derivative-free simplex method where the objective function is estimated using a linear interpolation at $2p + 1$ points. We use the version implemented in \texttt{SciPy} (\texttt{v1.5.2}) \cite{2020SciPy-NMeth}. \par
}

\medskip

{\setlength{\parindent}{0cm}
\underline{\textit{Powell}} \cite{Powell1964} - A derivative-free method formally known as ``Powell's conjugate direction method". Bi-directional line searches are carried out along orthogonal axes in parameter space from an initial point. The directions of new line searches and a new point are then determined by a linear combination of the optima of the previous line searches. We use the version implemented in \texttt{Nevergrad} (\texttt{v0.4.3.post9}) \cite{nevergrad} which itself draws from \texttt{SciPy}.
}

\medskip

{\setlength{\parindent}{0cm}
\underline{\textit{Sequential Quadratic Programming (SQP)}} - Specifically, the variation by Dieter Kraft \cite{Kraft1988}. The optimization of the objective function proceeds by solving individual quadratic programming subproblems. The solution to these quadratic models requires first and second derivatives of the objective function. The original Dieter Kraft code is interfaced using \texttt{Nevergrad} (\texttt{v0.4.3.post9}) \cite{nevergrad} which itself draws from \texttt{SciPy}. \par
}

\medskip

{\setlength{\parindent}{0cm}
\underline{\textit{Gradient Descent}} - The objective is optimized by stepping parameters in the direction of and proportional to the first derivative of the objective. We use the version implemented in \texttt{Pennylane} (\texttt{v0.19.0}) \cite{bergholm2020pennylane}. \par
}

\medskip

{\setlength{\parindent}{0cm}
\underline{\textit{Adam}} \cite{kingma2017adam} - A stochastic gradient descent algorithm utilizing machine learning techniques. We use the version implemented in \texttt{Pennylane} (\texttt{v0.19.0})  \par
}

\end{mdframed}

We now detail our benchmarking approach and justify it in the next paragraph. Using the problems instances defined in Section \ref{subsec:classicalopt}, we use each of the optimizers above to minimize $\langle \hat{H}_{\mathcal{C}} \rangle$ when evaluated using an exact statevector simulator and estimated using a finite number of 2048 shots both using \texttt{Qulacs} (\texttt{v0.3.0}) \cite{Suzuki2021}. From now onwards, a single evaulation of $\langle \hat{H}_{\mathcal{C}} \rangle$ by any method (i.e, by exact statevector simulation or from 2048 shots) will be referred to as a \textit{circuit evaluation}. For each simulator variation, we treat four constraint enforcement schemes (1) soft constraints using large $\alpha = 100$, (2) soft constraints using $\alpha = \alpha_{\text{min}}$, (3) hard constraints using Dicke state initialization and the complete graph mixer unitary ($|D_k^n \rangle$ \& $\hat{H}_K$) and (4) random Hamming state initialization and the ring mixer unitary ($|R_k^n \rangle$ \& $\hat{H}_R$). Now, for each of these constraint method/quantum circuit simulator variants, we minimize $\langle \hat{H}_{\mathcal{C}} \rangle$ using a fixed budget of 500 circuit evaluations, initializing $\gamma_p$ and $\beta_p$ randomly in the range $[0, 2\pi)$. If soft constraints are used, this process is repeated 50 times for each optimizer but if hard constraints are used we only repeat 10 times. For those optimizers requiring first or second order gradients of the objective function, we approximate them using finite differences. While there are in principle a number of hyperparameters which can be tuned for each of the optimizers, we choose to use the defaults which can be obtained using the supplied software package versions. From each run, we examine $\eta$ and $\mathcal{W}$.  \par

\begin{figure*}
    \centering
    \includegraphics[width=\linewidth]{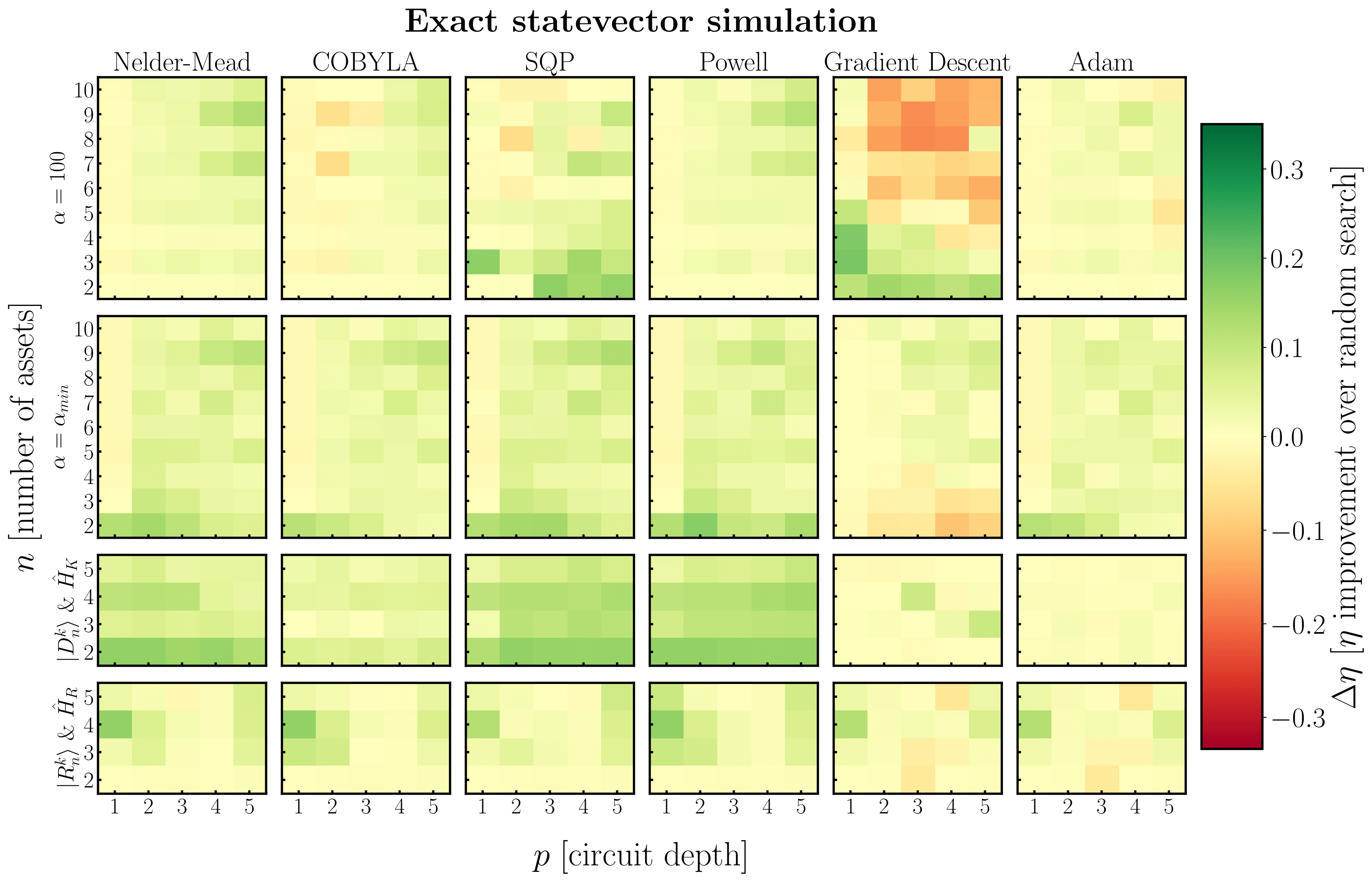}
    \caption{The best values of $\eta$ found after classical optimization (optimizers are organized into columns) compared to Random Search over all repeats at all considered $n$, $p$ and constraint enforcement schemes (organized into rows). Exact statevector simulator results are shown. Nelder-Mead, COBYLA and Powell are derivative-free optimizers, Gradient Descent and Adam require first derivatives of the objective while SQP requires first and second derivatives of the objective.}
    \label{fig:best_eta_sv}
\end{figure*}

\begin{figure*}
    \centering
    \includegraphics[width=\linewidth]{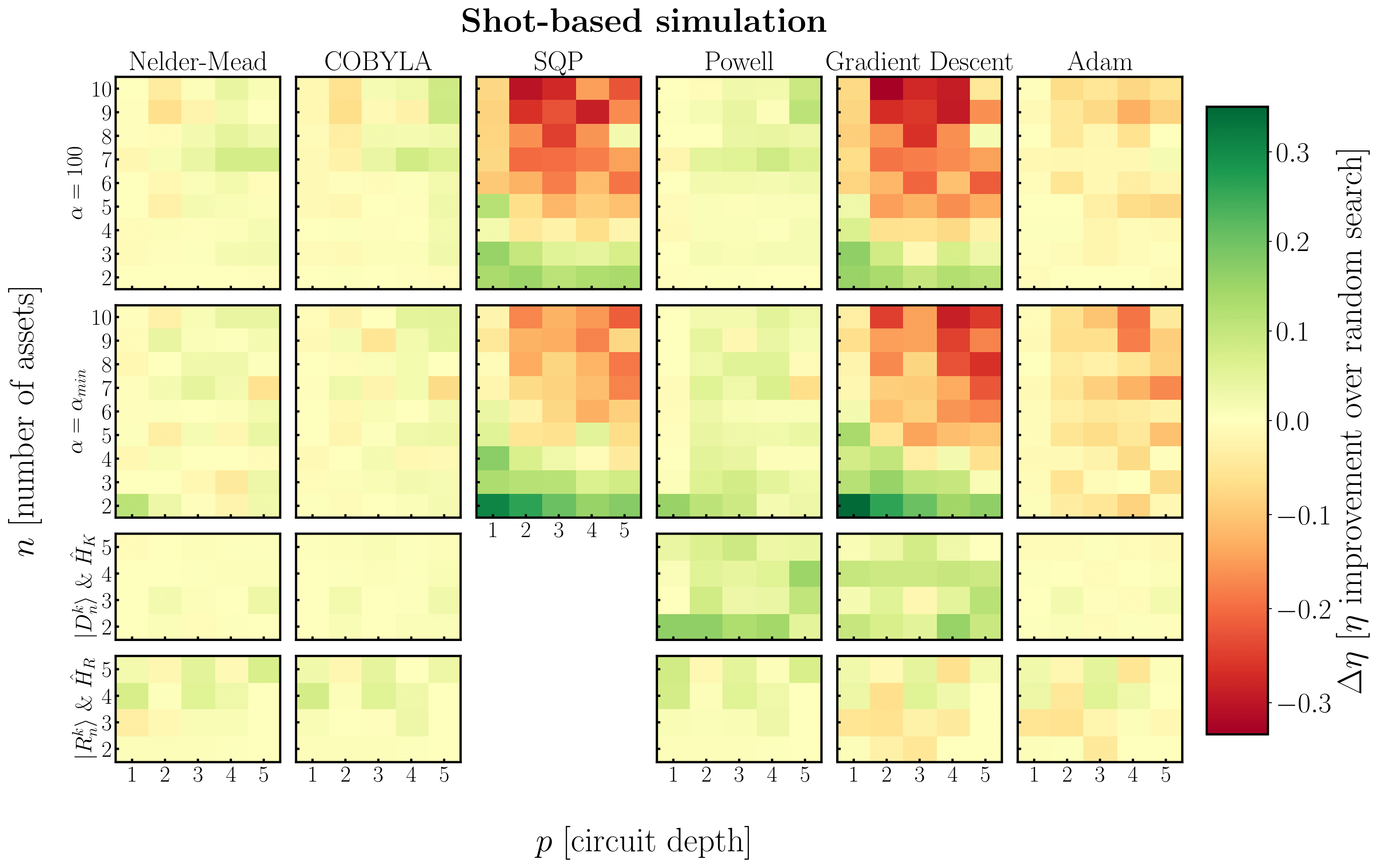}
    \caption{See the caption of Figure \ref{fig:best_eta_sv} but shot-based simulation of 2048 shots is used. Hard constraint data is omitted for SQP as is explained in the main text.}
    \label{fig:best_eta_shot}
\end{figure*}

The selection of optimizers we have chosen represents a broad sample of black box optimizers operating using different techniques. Some require evaluations of derivatives while some are derivative-free. For those requiring derivatives, SQP and Gradient Decent do not account for stochastic noise while Adam does. For the derivative-free optimizers, none are designed specifically for noisy objectives but compared with usual derivative methods, are known to perform well in the presence of noise \cite{bortz1998simplex, powell2007view}. With the exception of Random Search, all of the optimizers we treat are local minima finders. Random Search can in principle be considered a global minima finder should the bounds of the objective function match the bounds of the search. The possibility of local minima is what motivates repeating optimization runs at new random initial $\boldsymbol{\gamma}$ and $\boldsymbol{\beta}$. The different number of repeats among soft and hard constraint runs is motivated by initial runs and empirical analysis of cost landscapes showing a larger number of local minima for soft versus hard constraints. While it is possible that our choice of initial angles $\boldsymbol{\theta}_{init} \in [0, 2\pi)$ could influence the behaviour some optimizers to find minima local to these points, in our experiments, optimizers find minima far from these points (especially in $\gamma_p$ which extends to $\sim 100 \pi$). The choice of 500 circuit evaluations is to strike a balance between the likelihood of any optimizer finding good local minima and computational expense of the simulations. Also, since we wish our simulator results to maintain some transferability to what is possible on real QPUs, we must consider what the present state of the art is for reasonable quantum resources. That is, we contend that given the competition for access to a small number of QPUs which presently exist in this NISQ era and the means of accessing them (submitting jobs to a queue over the cloud), 500 circuit evaluations is at the upper limit of the reasonable time a user can be expected to wait for results of a single optimization. With further development of tools like \texttt{Qiskit Runtime} and \texttt{AWS Jobs} alongside increased accessibility to hardware, however, more circuit evaluations may soon become tolerable.

\subsubsection{Optimizer performance analysis \label{subsubsec:soln_qual_after_opt}}

This Section discusses the $\eta$ achieved given the approach presented in Section \ref{subsubsec:blackbox}. Most of the discussion references Figure \ref{fig:best_eta_sv} and \ref{fig:best_eta_shot} but we occasionally reference data in the \href{publisher.to.insert.link}{Supplemental Material}. A large portion of our analysis is centered around the comparison of the performance of different classical optimizers as compared with a Random Search. Indeed, one should expect a successful optimizer to perform better than simply randomly selecting $\boldsymbol{\gamma}$ and $\boldsymbol{\beta}$. Figure \ref{fig:best_eta_sv} and \ref{fig:best_eta_shot} show this comparison based upon the best $\eta$ found after all optimization repeats for exact statevector and shot-based simulation, respectively. \par 

Overall, we find that within the scope of our benchmarks, more often than not, derivative-free optimizers outperform derivative-requiring ones. This is true for exact statevector and shot-based simulation. This is with the notable exception of SQP paired with statevector simulation which performs similarly to the derivative-free methods. This observation is particularly noteworthy for hard constraints as working at a fixed Trotter step size makes $\langle \hat{H}_{\mathcal{C}} \rangle$ non differentiable (along the $\beta_p$ directions) which invalidates the key assumption of double-differentiability inherent to SQP \cite{Kraft1988}. Indeed, when doing the same experiments with shot-noise, the SQP algorithm stalls in-between circuit evaluations. As a consequence, we do not display SQP results for hard constraints on Figure \ref{fig:best_eta_shot}. In general, the solution quality gap between derivative-requiring and derivative-free optimizers widens when shot-based simulation is used. Although, a decrease in the performance of the derivative-free optimizers also observed. Much of this performance loss can be attributed to the ill definition of gradients on a stochastic objective when estimated using finite difference methods. That being said, Adam still shows notable performance decreases when shot-noise is present despite being more resilient to noise by design \cite{kingma2017adam}. We do note that performance losses are less pronounced than the other derivative-requiring optimizers which (especially at large $n$ and $p$) can perform much worse than random parameter selection. 

Among the different constraint methods, we can see that optimizers rarely perform worse than random search for hard constraints. From empirical observation of $p=1$ soft constraint cost landscapes versus hard constraint landscapes it can be seen that the latter is much less rough than the former. This simple fact is likely to impact the performance of most optimizers and increase the likelihood that the optimization terminates in local minima. When comparing between the two soft constraint variations, optimizers are more likely to perform better than random search when $\alpha = \alpha_{\text{min}}$. This can also be reasoned by the condition of the $p=1$ cost landscape. When $\alpha=100$, the cost landscape is dominated by a high frequency and high amplitude sinusoidal component. Above all, this means that far more local minima exist for optimizers to be trapped in. In the particular case of gradient descent paired with statevector simulation, we reason that poor performance is observed because the forward step in parameter space becomes large as driven by the large gradients of the high frequency and high amplitude sinusoidal component. The forward step then overshoots the minimum of the local basin of attraction thus projecting the optimizer towards a new (possibly shallower) basin of attraction. While this effect can likely be mitigated by tuning the hyperparameters of the optimizer, this is beyond the scope of this study. \par

So far, our discussion has been limited to the best solution quality achieved over all repeats. To obtain a full picture, however, it is necessary for some discussion about the mean and standard deviation over all runs. Figures demonstrating these metrics are presented in the \href{publisher.to.insert.link}{Supplemental Material}. When we examine the mean performance, it becomes much more likely that random parameter selection performs better than other optimizers. This especially true at low $n$ and low $p$ in soft constraint runs and even more true when shot noise is introduced. However, as both $n$ and $p$ increase, random searches become less effective. Indeed, at $n \geq 5$ and $p \geq 2$ $\eta(p)$ becomes a decreasing function for random search. For increasing $p$, this can be attributed simply to the ``curse of dimensionality" since the number of variational parameters scales with $2p$. For increasing $n$, the answer is less obvious. However, we do propose that the effect is a result of the increasing roughness of the cost landscape with $n$, which is in part related to the increasing total number of classical states able to contribute towards $\langle \hat{H}_{\mathcal{C}} \rangle$ and in part to do with the increasingly large gap in $\hat{H}_{\mathcal{C}}$ caused by the larger $n - B$ we choose to use at larger $n$. That is, unviable solutions are impacted by larger penalty factors which drive the troublesome high frequency and amplitude sinusoidal component in $\langle \hat{H}_{\mathcal{C}} \rangle$. \par

Should we now examine the standard deviation of $\eta$ across all runs (a measure of reliability, be that reliably good or bad), it becomes clear that the spread is in most cases much wider than random search. In the particular case of $p = 1$ in soft constraints, the spread is particularly large for all optimizers at some $n$ implying that a large number of local minima are present at minimal circuit depth. Th standard deviations do need to be understood with the context of the mean. Indeed, gradient descent is poor at high $n$ and $p$ on average, but the standard deviation is small (it is reliably poor). When examining the optimizer trajectories for Gradient Descent, even at 500 circuit evaluations, we often observe the non-convergence of the optimizer which is likely cause of the large standard deviation. In constrast, Adam is poor on average and has a large standard deviation (unreliably poor). For the derivative free optimizers which performed relatively well on average, we see that the standard deviations are large meaning that while they perform well sometimes, this is cannot be counted upon given any randomized cold-start. \par

To summarize the results of this Section, within the parameters of our simulations, derivative-free optimizers are most likely to out-perform derivative-requiring optimizers on the basis of (i) the best $\eta$ (ii) the mean $\eta$ and (iii) the standard deviation of $\eta$ over a number of randomly initialized QAOA runs compared with random search. For (ii) \& (iii), random searches often perform better than other optimizers, especially for shot-based simulations using soft constraints. We reason that this is because of the large number of local minima in the soft constraint cost landscape and the limited tolerance to noise of the optimizers tested. It is possible, however, that the performance of the more noise-tolerant Adam optimizer could improve with a larger number of circuit evaluations. Overall, hard constraint ans\"{a}tze were easier to optimize than their soft constraint counterparts as demonstrated by superior performance in (i-iii). Our results support current trends in the literature for classical optimizers in the QAOA and more general variational circuits. That is, we can view the most successful runs as being the product of good initial angles. Since the large spread in $\eta$ suggests that random initialization is not a good approach (as has been suggested in other works \cite{Farhi2014, Zhou2020}), we must look to other angle initialization schemes. Some recent approaches include: encoding the solution of the continuous relaxation of the problem into the initial state \cite{Egger2021}, using a quantum annealing inspired initialization \cite{Sack2021}, initializing the $p+1$ ans\"{a}tz with the level $p$ angles \cite{Zhou2020}, using optimal anz\"{a}tze from similar problems \cite{Shaydulin2019} and machine learning approaches \cite{Khairy2020, alam2020accelerating}. To further mitigate the influence of local minima, various techniques have been suggested including: standard basin hopping, coupling the cost landscape with a classical neural network \cite{riveradean2021avoiding} or using the quantum natural gradient \cite{Stokes2020quantumnatural, Wierichs2020}.

\subsubsection{Comparing constraint enforcement schemes and problem size scaling \label{subsubsec:soln_qual_constraints}}

\begin{figure}
    \centering
    \includegraphics[width=\linewidth]{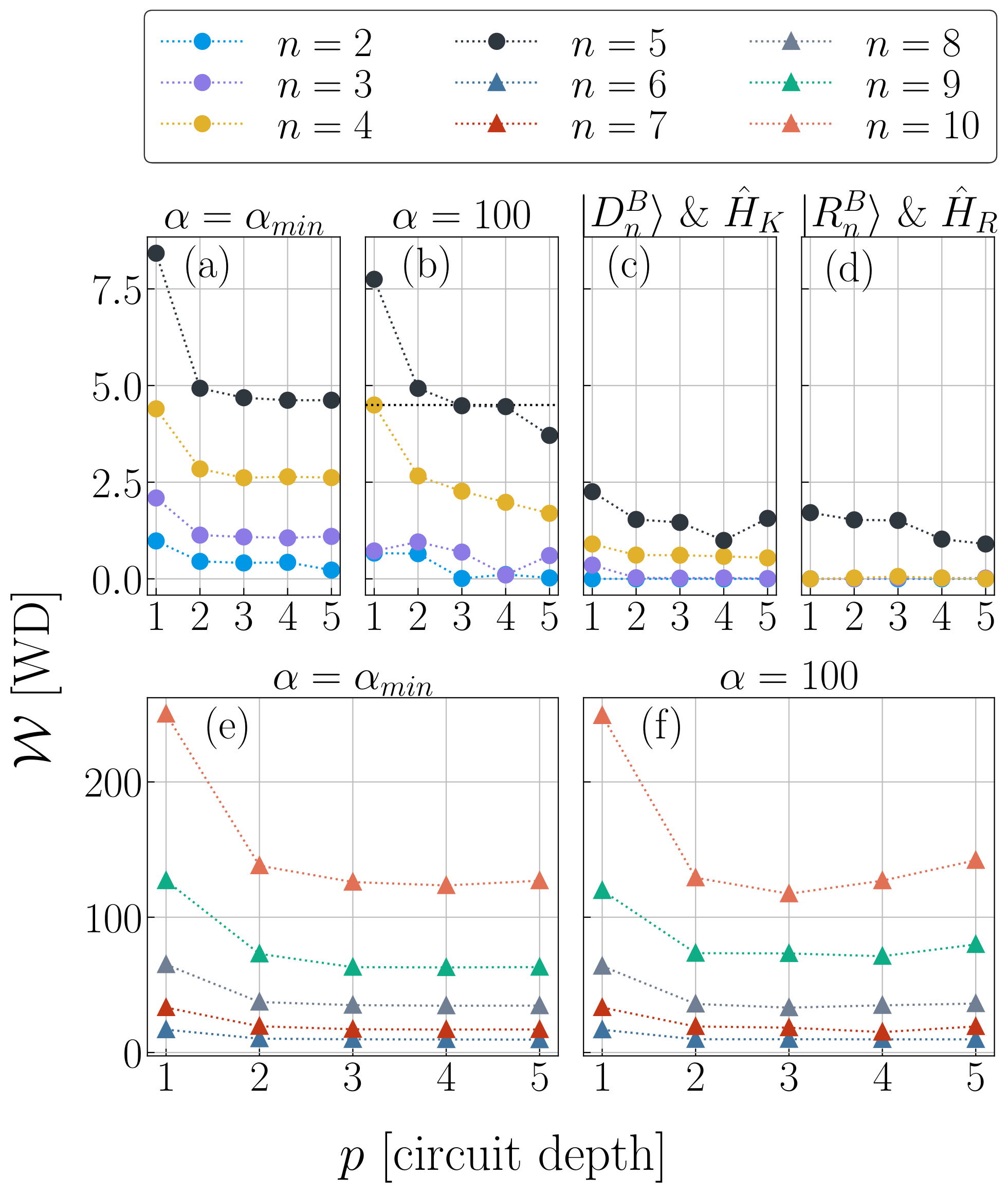}
    \caption{$\mathcal{W}(p)$ using the best $\boldsymbol{\gamma}$ and $\boldsymbol{\beta}$ from all simulator runs. Decreasing $\mathcal{W}(p)$ indicates that an increasing amount of probability is transported towards better solutions. (a-d) compares $n=1$:$5$ for all constraint methods and (e-f) compares soft constraint approaches for $n=6$:$10$. The dotted black horizontal line on (b) indicates the worst case ($p=1$) $\mathcal{W}$ at $n=4$ and its intersection with $n=5$ around $p=3$. Such a comparison can be used to measure how quantum computational effort scales with problem size.}
    \label{fig:wdconstraints}
\end{figure}

Should we examine best ans\"{a}tze from all runs, we can make some comparisons regarding the quality of solutions achieved by the different constraint enforcement schemes. Such a comparison is made in Figure \ref{fig:wdconstraints}. Naively, we see from Figure \hyperref[fig:wdconstraints]{8(d)} that $|R_n^B \rangle$ \& $\hat{H}_{R}$ produces the best quality solutions overall. However this is only because the random state $|R_n^B \rangle$ is often prepared as $|x^* \rangle$ in the best ans\"{a}tz which is likely to occur for small $n$ given the number of repeats. This is of course not scalable; a fact already observable by $n=5$ where $\mathcal{W}$ deviates from 0 becoming comparable with $|D_n^B \rangle$ \& $\hat{H}_{K}$ [Figure \hyperref[fig:NCWDexplained]{8(c)}]. For the treated range of $p$, we only observe comparable $\mathcal{W}$ among all constraint approaches for $n=2$. $\mathcal{W} \approx 0$ is achieved for all $p$ in hard constraints [Figure \hyperref[fig:NCWDexplained]{8(c-d)}] while this is only observed at higher $p=5$ for $\alpha = \alpha_{min}$ [Figure \hyperref[fig:NCWDexplained]{8(a)}] and $p=3$ for $\alpha = 100$ [Figure \hyperref[fig:NCWDexplained]{8(b)}]. Broadly, at $n \lessapprox 5$, we observe better performance with increasing $p$ in soft constraints for $\alpha=100$ than for $\alpha = \alpha_{\text{min}}$. This trend appears to reverse for $n = 9$ and $10$, even showing a noticeably increasing $\mathcal{W}$ for $p > 3$. For all soft constraint approaches and for $n > 3$ in $|D_n^B \rangle$ \& $\hat{H}_{K}$, we observe the best reduction of $\mathcal{W}$ when $p$ increases from $1$ to $2$. Beyond this, we observe a much more gradual reduction, in a lot of cases flattening to only marginal improvements with $p$. \par

We now make some crucial comments about how $\mathcal{W}$ scales with problem size within a given constraint enforcement scheme. In soft constraints and for $n > 3$ in $|D_n^B \rangle$ \& $\hat{H}_{K}$, at fixed $p$, we observe an exponential scaling in $\mathcal{W}$ with $n$: $\mathcal{W} \propto 2^n$. While this might initially appear troublesome for the success of the QAOA, since we observe strong reductions in $\mathcal{W}$ with only small integer increases in $p$, within the bounds of our simulations, we can observe the overcoming of this exponential wall. This is most obvious when we observe $\mathcal{W}(p)$ at some $n$ reducing below the worst case $\mathcal{W}$ (most like at $p=1$) at $n-1$. This is occurs many times on Figure \ref{fig:wdconstraints} with one representative instance indicated on Figure \hyperref[fig:wdconstraints]{8(b)}. Here we observe $\mathcal{W}(p)$ for $n = 5$ reduce below the worst case $\mathcal{W}$ for $n=4$ when $p \geq 3$. We suggest that a wider study examining these points of intersection could prove vital for establishing heuristic scaling behaviours in problem size for a wide class of combinatorial optimization problems in the QAOA.        

\section{Solution quality using real gate-model QPUs \label{sec:realqpu}}

To demonstrate the utility of $\eta$ for use in application specific benchmarking, we now use it to measure the performance of QAOA-based PO on a selection of gate-model QPUs. Our goal is to evaluate $\eta(p)$ among different cloud-accessible QPUs to obtain a \textit{naive} notion of the success of QAOA-based PO on NISQ-era devices. We label our approach \textit{naive} as our goal is not to extract the best possible performance from each QPU but rather to provide a baseline as given by the defaults of each provider. Crucially, this means that the results presented in the following Sections are free from any micro-optimizations which could include: control over circuit transpilation optimization levels \cite{Harrigan2021, weidenfeller2022scaling}, pulse level control of gates \cite{Earnest2021} and others. Our results are also free of any error mitigation techniques which are known to improve performance of the QAOA \cite{barron2020measurement, Bravyi2021, Harrigan2021, weidenfeller2022scaling}. This approach is motivated by the fact that (presently) different providers offer differing levels of support for user defined micro-optimization. For example, as of the time of writing, it is possible for some providers to locally control circuit transpilation optimization levels while for others this is completed by the provider after circuits are sent over the cloud. Indeed, the exact circuit being executed on some cloud QPUs (after cloud-side transpilation into the native gates) are not yet accessible. \par

Our benchmarks are based upon a limited re-optimization of the best ans\"{a}tze determined by quantum circuit simulators for the real stock market problem instances of the previous Section. Specifically, limited to 10 circuit evaluations, we \textit{warm start} a COBYLA optimization of $\langle \hat{H}_C \rangle$ with the optimal angles under the execution of an ideal QPU ($\boldsymbol{\gamma}^{\text{ideal}}$ and $\boldsymbol{\beta}^{\text{ideal}}$) using 2048 shots per circuit evaluation. The motivation for this approach is to allow any small mismatches (understandable just on the basis of imperfect gate fidelity) in $\boldsymbol{\gamma}$ and $\boldsymbol{\beta}$ between the ideal and experimental cost landscapes to be corrected. Our choice of using just 10 circuit evaluations is discussed in the \href{publisher.to.insert.link}{Supplemental Material} as compared to a full 500 circuit evaluations as were used in the previous Section. Broadly, we find 10 circuit evaulations sufficient and further evaluations do not statistically improve performance. Following this approach, we benchmark QPUs made available by IBM \cite{ibmq2021}, Rigetti \cite{rigetti2021} and IonQ \cite{ionq2021}. IBM QPUs were accessed directly through the provider while Rigetti and IonQ QPUs were accessed through AWS Braket. From IBM, we use \texttt{ibmq\_manila}, \texttt{ibmq\_bogota}, \texttt{ibmq\_quito}, \texttt{ibmq\_belem} and \texttt{ibmq\_lima}. From Rigetti, we use \texttt{Aspen-10} and from IonQ we use the 11 qubit machine (abbreviated to \texttt{11-Q} in the rest of this work). These choices of QPU span two different hardware paridigms (superconducting vs trapped ion qubits) and several different qubit connectivity topologies. Relevant details for each of these QPUs are tabulated in the \href{publisher.to.insert.link}{Supplemental Material}. Our benchmarks are in the range $2 \leq n \leq 5$, $1 \leq p \leq 5$ for $\alpha = \alpha_{\text{min}}$ soft constraints, $|D_n^B \rangle$ \& $\hat{H}_{K}$ and $|R_n^B \rangle$ \& $\hat{H}_{R}$. Indeed, beyond  the given range of $n$, $\alpha = \alpha_{\text{min}}$ soft constraints (the method with the shallowest depth) produces a distribution of solutions indistinguishable from random ($\eta \approx 0.5$) for all QPUs; tested at $n=6$, $p=1$. The results of these benchmarks are discussed in Section \ref{subsec:perform_vs_depth_real}. To investigate reproducibility among the different QPUs, $\alpha = \alpha_{\text{min}}$ soft constraint runs were repeated 5 times (with the exception of IonQ's \texttt{11-Q} machine which we repeated twice). This is discussed in Section \ref{subsec:variability}. Soft constraints using $\alpha = 100$ were excluded from QPU benchmarks because of the possibility of errors incurred transferring the simulated optimal angles over to the real QPUs. This is following the findings of \cite{ozaeta2021expectation} which suggested that narrow valleys in $\langle \hat{H}_{\mathcal{C}} \rangle$ (like are seen for $\alpha = 100$) could be missed on NISQ hardware.

\subsection{Solution quality and fidelity versus depth \label{subsec:perform_vs_depth_real}}

\begin{figure}
    \centering
    \includegraphics[width=\linewidth]{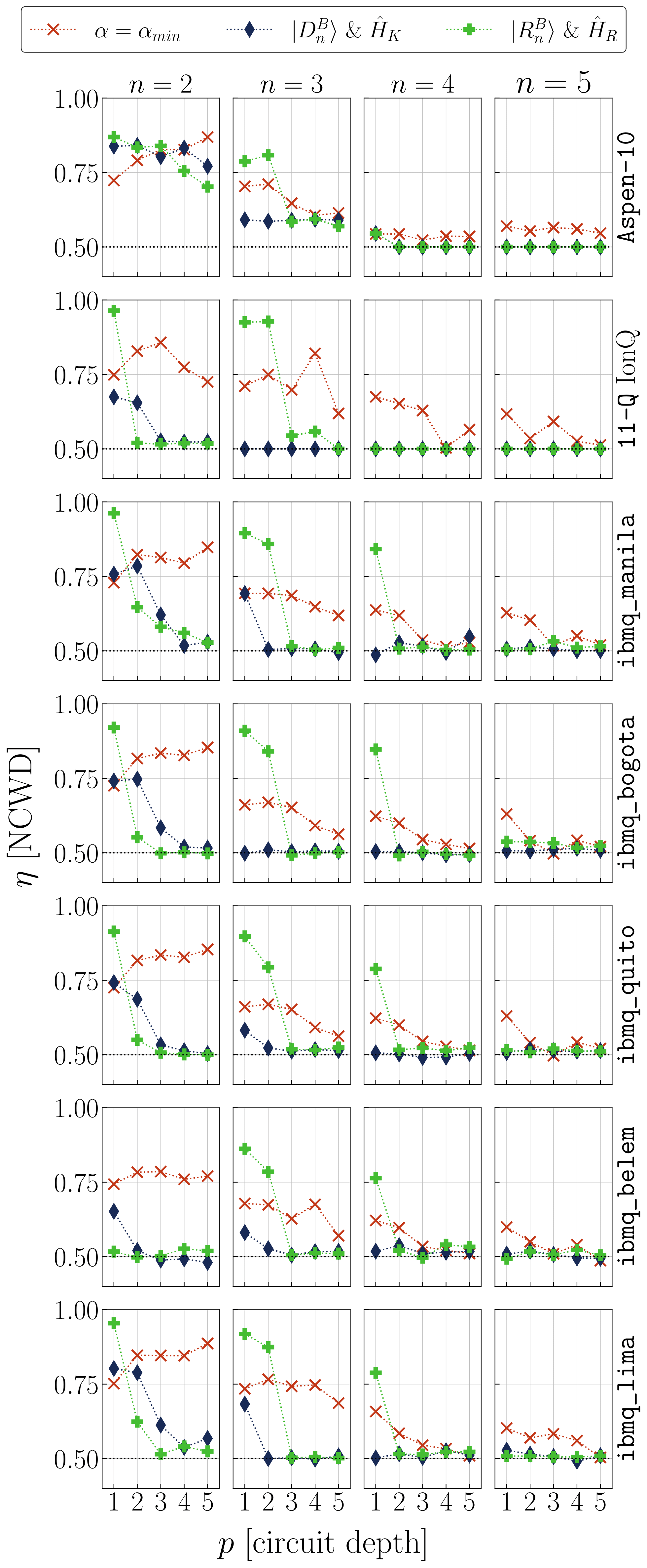}
    \caption{$\eta(p)$ for a selection of gate-model QPUs for $2 \leq n \leq 5$, $1 \leq p \leq 5$ and three constraint enforcement schemes: $\alpha = \alpha_{\text{min}}$, $|D_n^B \rangle$ \& $\hat{H}_{K}$ and $|R_n^B \rangle$ \& $\hat{H}_{R}$. Soft constraint $\alpha = \alpha_{\text{min}}$ runs use the highest $\eta$ achieved over 5 repeats (2 for \texttt{11-Q} IonQ) while hard constraint runs were not repeated. Decreasing/low $\eta(p)$ indicates the presense of noise.}
    \label{fig:eta_func_depth_qpu}
\end{figure}

\begin{figure*}
    \centering
    \includegraphics[width=\linewidth]{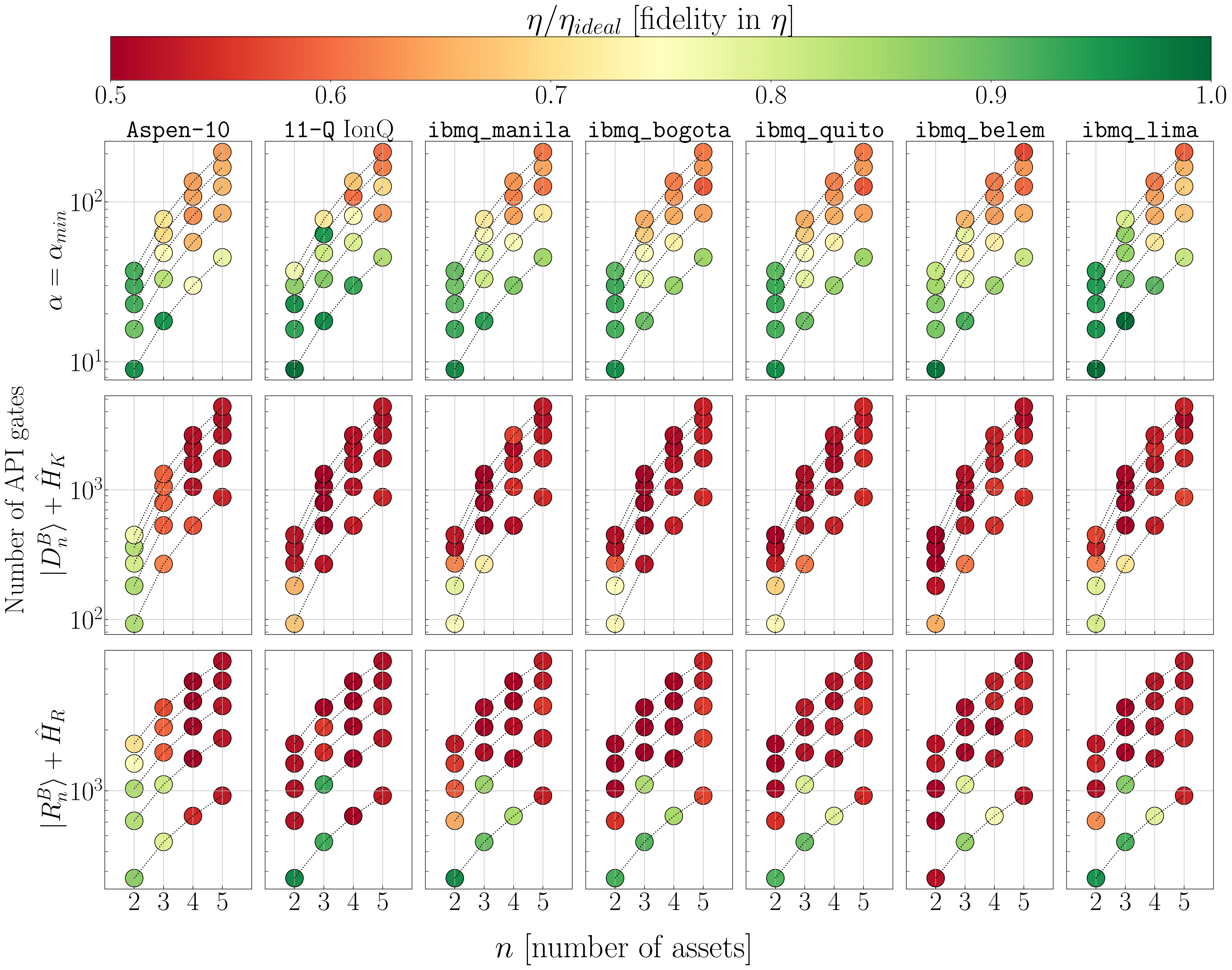}
    \caption{The fidelity in $\eta$ ($\eta/\eta_{\text{ideal}}$) for a selection of gate-model QPUs (organized into columns) for $2 \leq n \leq 5$, $1 \leq p \leq 5$ and three constraint enforcement schemes: $\alpha = \alpha_{\text{min}}$, $|D_n^B \rangle$ \& $\hat{H}_{K}$ and $|R_n^B \rangle$ \& $\hat{H}_{R}$ (organized into rows). Soft constraint $\alpha = \alpha_{\text{min}}$ runs use the highest $\eta$ achieved over 5 repeats (2 for \texttt{11-Q} IonQ) while hard constraint runs were not repeated. Markers connected by a black dotted line share a common QAOA circuit depth $p$. The logarithmic scale $y$-axis is given as the number of gates used in the API circuit (explained in the text). This is distinct from the number of native gates which are actually executed on the QPU.}
    \label{fig:eta_fidelity}
\end{figure*}

We discuss our benchmarks with reference to Figure \ref{fig:eta_func_depth_qpu} and \ref{fig:eta_fidelity}. The former shows raw $\eta(p)$ performance for each QPU and the latter shows the fidelity of $\eta$ as compared to the $\eta$ obtained by the ideal simulator ($\eta_{\text{ideal}}$) in Section \ref{subsubsec:soln_qual_after_opt} ($\eta/\eta_{\text{ideal}}$). The fidelity is given as a function of the number of application programming interface (API) gates. This is the number of gates in $G_{\text{API}} =$ \{$H$, $CNOT$, $RZ(\theta)$, $RX(\theta)$, $RY(\theta)$\} used to construct the QAOA circuits at the API level. As mentioned in the previous Section, this is distinct from the number of native gates which are executed on the machine, so, our results also reflect the performance of the default settings of the vendor-specific transpilers. The native gates of each QPU are given in the \href{publisher.to.insert.link}{Supplemental Material}. For hard constraint runs, our choice of working at a fixed Trotter time step means that the number of API gates is proportional to the present value of $\boldsymbol{\beta}$ in the optimization. Subsequently, the number of API gates are given as the average number of gates required within the bounds of $\boldsymbol{\beta}$ for the given mixing Hamiltonian. For $\hat{H}_K$, $\boldsymbol{\beta}$ has definite bounds but for some values of $n$, $\hat{H}_R$ gives rise to incommensurate bounds for $\boldsymbol{\beta}$. However, from observation, we find that runs using $\hat{H}_R$ do not escape $0 \leq \beta_p \leq 2\pi$ so we use these bounds in the average. The exact number of API gates used for the angles producing the largest $\eta$ (exactly at the points shown on Figure \ref{fig:eta_fidelity}) are given in the \href{publisher.to.insert.link}{Supplemental Material}.

Among the different constraint methods, the most consistently well performing approach on all QPUs is $\alpha = \alpha_{\text{min}}$ soft constraints. As shown in Figure \ref{fig:eta_func_depth_qpu}, at $n=2$, it is possible to observe a monotonic increase in $\eta(p)$ for soft constraints up to $p=5$ although this trend does vary between QPUs. At $n=3$, $\eta(p)$ ceases to be monotonic for all machines with the largest $\eta$ appearing at either $p=2$ or 4. At $n=4$ and $5$, we no longer observe improvement with $p$ for any QPU but see results better than random guessing ($\eta > 0.5$) all the way to $p=5$ at $n=5$. Since to the author's knowledge, no real QPU results have been reported for QAOA-based MVPO, we measure these results against the QAOA solution to other QUBO problems. This work shows an improvement over the results of \cite{alam2019analysis} where a related benchmark on an unconstrained 3-node complete graph max-cut problem showed monotonically decreasing performance with $p$ using a previous generation IBM QPU (\texttt{IBMQX4}). Indeed, since the Ising model formulation of max-cut involves only quadratic terms, our performance gains are even larger since the MVPO Ising model has linear terms which require additional single qubit gates to be implemented in the cost unitary. These improvements should also be understood in the context of the first observation of increasing success probability with QAOA circuit depth on hardware \cite{Bengtsson2020}. That is, just three years prior to this work, increased performance at $p=2$ for two superconducting transmon qubits was landmark. Our benchmarks now show that such an observation is commonplace among different QPU paridigms at larger $n$ and $p$. \par

We also remind the reader that our benchmarks are free from micro-optimization so its plausible that performance could still be increased. Indeed, recent results from Google \texttt{Sycamore} \cite{Harrigan2021} showed relatively $n$-independent performance up to 23 qubits to solve a subset of graph problems optimized for and directly mapable to the hardware grid. For these problems, they observed a peak in performance at $p=3$ when $n > 10$ \cite{Harrigan2021} with the caveat that the type of graph problems studied are \textit{efficiently solvable with classical algorithms}. For problems which were not optimized for the qubit topology, and like discrete MVPO, \textit{not} efficiently classically solvable (max-cut and the Sherrington Kirkpatrick model; see the supplement of \cite{Harrigan2021}), increasing performance with $p$ was not observed, unlike in our results where it is. \par

Hard constraint runs are generally less successful than soft constraints on all QPUs. For both $|D_n^B \rangle$ \& $\hat{H}_{K}$ and $|R_n^B \rangle$ \& $\hat{H}_{R}$, $\eta(p)$ is broadly a decreasing function (although there are exceptions); understandable from the almost order of magnitude larger number of API gates required in their implementation (See y-scale of Figure \ref{fig:eta_fidelity}). There are, however, specific instances where $\eta$ is either comparable with or out performs soft constraints. For many QPUs, this is observed at low $n$ and $p$, especially for $|R_n^B \rangle$ \& $\hat{H}_{R}$ at $n=2$ and 3. Here, we see $\eta \approx 1$ although this can be partially attributed to the small number of gates required for the to implement the Trotterized mixing unitary at small $\mathcal{\beta}$. Beyond $n=4$, $p=1$ all QPUs tend to produce $\eta \approx 0.5$ for both hard constraint approaches. However, for Rigetti's \texttt{Aspen-10} and IonQ's \texttt{11-Q} machines, we were unable to complete the full set of hard constraint benchmarks at high depth because of time-outs in cloud-side circuit transpilation. Specifically, for \texttt{Aspen-10}, this occurs beyond $n = 4$, $p = 1$ for $|D_n^B \rangle$ \& $\hat{H}_{K}$ and $n = 4$, $p = 2$ for $|R_n^B \rangle$ \& $\hat{H}_{R}$. For \texttt{11-Q} this occurs beyond $n = 2$, $p = 5$ for $|D_n^B \rangle$ \& $\hat{H}_{K}$ and $n = 3$, $p = 4$ for $|R_n^B \rangle$ \& $\hat{H}_{R}$. For the instances where transpilation timed out, $\eta$ is set to $0.5$ on Figure \ref{fig:eta_func_depth_qpu} and \ref{fig:eta_fidelity}. An important observation is the small increase in $\eta(p)$ from $p = 1$ to $2$ observed for \texttt{ibmq\_manila}, \texttt{ibmq\_bogota} and \texttt{Aspen-10} for $|D_n^B \rangle$ \& $\hat{H}_{K}$. We can also see this happen for some instances for $|R_n^B \rangle$ \& $\hat{H}_{R}$ but for the $n$ this is observed, the ideal $\eta$ from simulation is exactly 1 for all $p$. I.e, the increasing $\eta$(p) on hardware does no correspond with $\eta$(p) in ideal execution. The results shown here the first implementations of hard constraint QAOA \cite{Hadfield2019} on real hardware and the first to show increasing solution quality with $p$. This is a result of equal calibre to the first observations of increasing success probability on a real QPU \cite{Bengtsson2020} for the original formulation of QAOA \cite{Farhi2014}. On the other hand, Figure \ref{fig:eta_fidelity} does show that there is much more room for improvement as although we do observe an increase, the fidelity to the ideal result remains low. \par 

Having now discussed some general trends, we can examine in more detail the results from some individual QPUs.  On \texttt{Aspen-10}, hard constraint runs execute rather successfully, producing a distribution of solutions better than random throughout all $p$ at $n=3$. At $n=2$ the QPU produces comparable results to soft constraints. One reason for this success could be related to the native implementation of $XY$ gates on the QPU \cite{Abrams2020}. That is, since $\hat{H}_K$ and $\hat{H}_R$ are themselves $XY$ operators, the unitaries they generate need not be decomposed into other gates on \texttt{Aspen-10}. The overall performance of soft constraints on \texttt{11-Q} is robust. Usually, $\eta(p)$ is non-monotonic but features large peaks alluding to the presence of significant coherent noise. Although the quantum volume (QV) of this machine has not been formally measured, a previous work estimated it at 64 \cite{lubinski2021applicationoriented}. This fact alongside the qubit graph topology (a complete graph) makes \texttt{11-Q} (i) the machine with the largest QV benchmarked in this study (although there have been no volumetric benchmarks of \texttt{Aspen-10}) and (ii) the only machine supporting native connectivity of all qubits in the range $n = 2$:$5$. Since MVPO is intrinsically a problem on a complete graph (Figure \ref{fig:graph_mvpo}) this is the likely source of the performance on \texttt{11-Q}. That being said, these factors do not clarify why this performance then degrades for $|D_n^B \rangle$ \& $\hat{H}_{K}$. \par

Indeed, qubit topology and QV do not in general predict the performance of the QAOA on even the most directly comparable of machines; the IBM QPUs. These machines share the same native gate set and have all been volumetrically benchmarked by IBM. If we were to go by QV alone, \texttt{ibmq\_manila} or \texttt{ibmq\_bogota} should produce the best overall performance. They do not. If we were to then take into account their 1D-chain qubit topologies (at most 3 connected qubits), it could be argued that it should be either \texttt{ibmq\_quito} or \texttt{ibmq\_belem} producing the best results for MVPO as their T-shape topology (at most 4 connected qubits) could make up for their lower QV of 16. This is also not true. Remarkably, among the IBM machines, the QPU with the lowest QV of 8 - \texttt{ibmq\_lima} - produces the best $\eta$ on average for soft constraints. We must then conclude that current generalized benchmarks do not appear predictive. Indeed, our observations could be explained by a recent work \cite{Proctor2021} showing that QPUs have measurably different noise characteristics for random circuits (QV benchmarking) than they do for structured ones (like the QAOA) suggesting that future benchmarking approaches need to account for both. Furthermore, it is reasonable to expect that quantum circuit structures are on a sliding scale between randomized and structured rather than strictly one or the other. Different applications will sit on different positions on this scale meaning that QPUs could give rise to characteristically differently performance profiles for different applications thus highlighting the need for application specific benchmarks.

\subsection{Variability in solution quality \label{subsec:variability}}

\begin{figure}
    \centering
    \includegraphics[width=\linewidth]{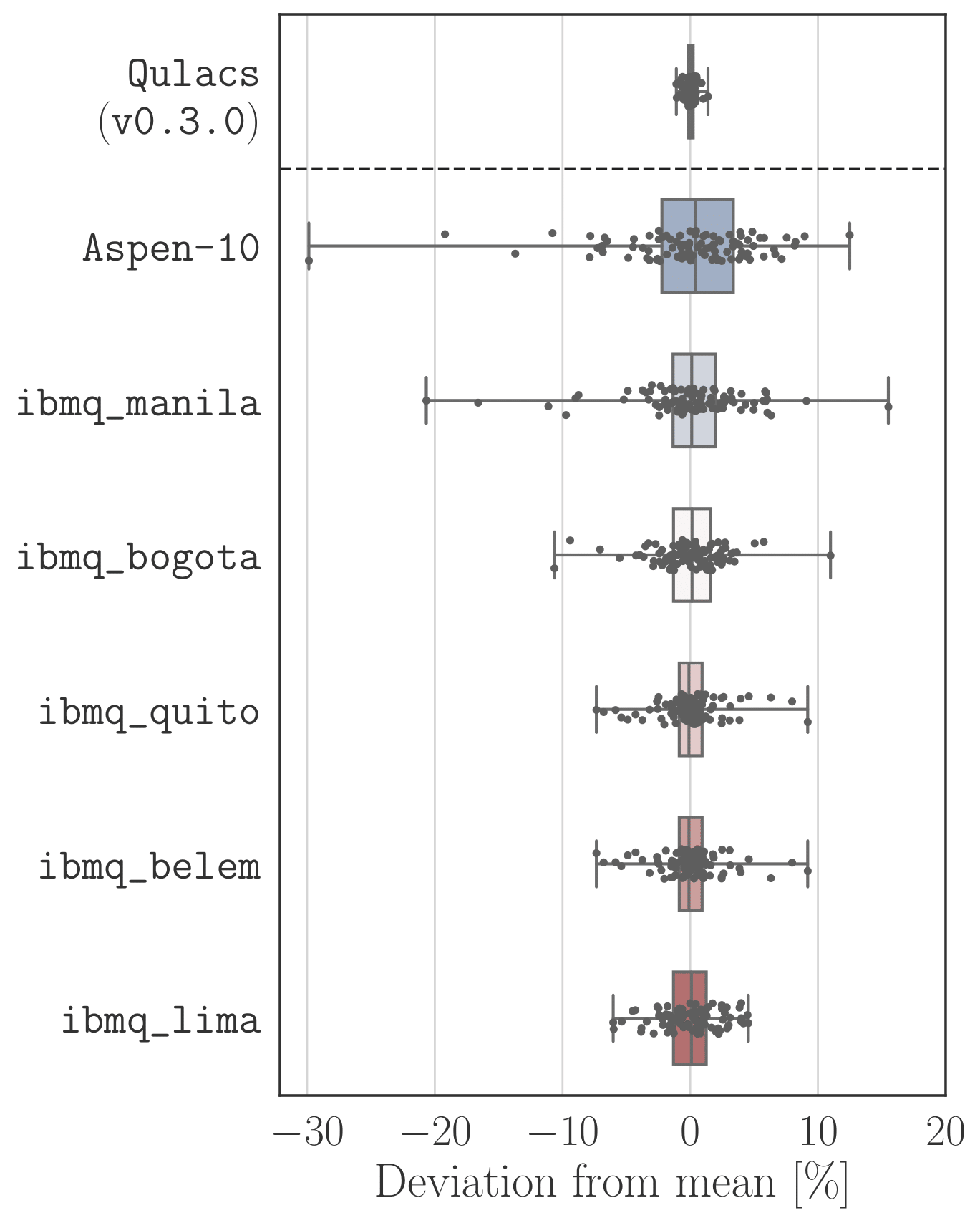}
    \caption{The percentage deviation of $\mu$ at some $n$ and $p$ from the mean value of $\eta$ ($\bar{\eta}$) over all 5 repeats for the same $n$ and $p$. Soft constraints using $\alpha = \alpha_{\min}$ were used. Shot-based simulation results (using \texttt{Qulacs} [\texttt{v0.3.0}] \cite{Suzuki2021}) are included above the black dashed line to in order to compare the shot-based stochastic noise against the observed QPU variability. We see that variability can be large and well exceeds variability from shot-based stochastic noise.}
    \label{fig:eta_variability}
\end{figure}

Soft constraint runs were repeated 5 times for each QPU apart from for IonQ's \texttt{11-Q} which were repeated twice. From these repeats we can estimate the variability of each QPU. For a given value of $n$ and $p$, our approach is to calculate the deviation of $\eta_{init}$ (the initial value of $\eta$ in the QPU re-optimization; $\boldsymbol{\gamma}$ and $\boldsymbol{\beta}$ are constant) for individual repeats from the mean of all 5 repeats. We do this for all treated values of $n$ and $p$ and display the mean deviations on Figure \ref{fig:eta_variability}. This Figure does not include \texttt{11-Q} results since the smaller number of repeats means they are not of equivalent statistical significance to the others. We do note, however, that we observed deviations of up to $19\%$ from the midpoint of the two repeats for this QPU. It is clear from Figure \ref{fig:eta_variability} that the performance of QAOA-based MVPO is strongly variable between repeated identical runs at a level much larger than the stochastic noise from the finite number of shots (see the uppermost boxplot on Figure \ref{fig:eta_variability}). A recent work also made this observation for different applications \cite{Blinov2021}, speculating that this phenomenon could be the result of transpilers choosing different qubit assignments between runs, calibration slippages or degradation of the entire system over time. To add to this discussion, we propose that part of this variability could be explained by the time-varying nature of qubit coherence times as shown in some recent works \cite{EtxezarretaMartinez2022, EtxezarretaMartinez2021, Burnett2019, Klimov2018}. We suggest that the observed variation is most likely a combination of all of these factors as well as others which are presently unknown to us. Whatever the source, it is clear from our results that any application specific benchmarks for any QPU should also be judged on the basis of their variability which in itself is a different and new benchmark.

\section{Conclusions \label{sec:conclusions}}

With a focus on MVPO, we have demonstrated that the performance of the QAOA can be measured using the concept of WDs \cite{radha2021quantum} and have presented a formalism for doing so; valid for any combinatorial optimization algorithm whose output is a probability distribution of solutions to a decision problem. Our metric, $\eta$, allows the QAOA to be intuitively viewed as a transporter of probability, able to judge the success of the QAOA in a way which is agnostic to the numerical details of different problem instances and constraint enforcement schemes. \par

Using this metric, at $p=1$, we identified that $\eta$ evolves with $B/n$ in QAOA-based MVPO, independent of the method used to enforce constraints. We found that problem instances with $B/n$ closest to the extrema (1 or $n-1$) allowed for the QAOA to produce the highest quality solutions while those closest to $B/n = 0.5$ showed the lowest quality solutions for the same computational effort. Remarkably, this directly corresponds with the evolution of the binomial coefficient $nCB$; the number of viable solutions. While we observe this phenomenon specifically for MVPO, we suggest that similar phenomena may exist for a wider class of constrained problems where solution quality (as determined by $\eta$) will correspond with the number of viable solutions. This finding brings a greater understanding of the possible bridges between notions of classical and quantum difficulty which have been debated in recent works \cite{Akshay2020, zhang2021quantum}. 

Moving beyond $p=1$, we used $\eta$ to benchmark the success of a variety of different black-box classical optimizers for minimizing the expectation value of the cost Hamiltonian $\langle \hat{H}_{\mathcal{C}} \rangle$ with exact statevector and shot-based quantum circuit simulators. In general, in the ranges $2 \leq n \leq 10$, $1 \leq p \leq 5$, given a fixed number of circuit evaluations, derivative-free optimizers outperformed derivative-requiring ones. Hard constraint circuits were found to be easier to optimize than soft constraints; a fact we attribute to the ill condition of the cost landscape for soft constraint problems. When a finite number of shots are used, the incurred stochastic noise significantly degrades the performance of all optimizers which for most cases makes performance comparable to a random search over the parameter space. Given the large spread of $\eta$ over different cold-start repeats of the optimization, it is clear that many local minima exist in the cost landscape, especially for soft constraints. This supports current research directions related to global minimization \cite{riveradean2021avoiding} and warm-starting \cite{Egger2021, Sack2021}. Using the WD ($\mathcal{W}$), we were able to provide insights into how quantum computational difficulty scales with increasing $n$ and $p$ having identified the existence of critical $p$-values where the solution quality of $n$ and $n \pm 1$ intersect. We suggest that the $\mathcal{W}$ metric could be used to study the same scaling behaviours for any combinatorial optimization problem solved using the QAOA. \par

Given a selection of gate-model QPUs, we used $\eta$ as an application specific benchmark of their performance for QAOA-based MVPO. In the process of doing so, we showed the first deployment of hard constraint QAOA \cite{Hadfield2019} to real QPUs which in some instances led to the observation of increasing solution quality with depth. Indeed, for the canonical QAOA \cite{Farhi2014}, it is now commonplace to observe increasing solution quality with depth on real QPUs. In the best cases, this is observed for $n = 2$ and $3$ up to $p=5$ and $p=4$ respectively. Measured against the results of approximately comparable previous works \cite{alam2019analysis, Bengtsson2020}, we have observed the improvement of gate-model QPUs over time for combinatorial optimization problems. Among the IBM QPUs, we observed a lower QV QPU producing higher quality solutions than QPUs with higher QV and the same qubit topology. This highlighted the need for application specific benchmarking as we observe that general benchmarking metrics are not predictive of application performance. Furthermore, we observed that the quality of solutions produced by all QPUs varied at a level much larger than the stochastic noise from the finite number of shots. While the origin of this variance is likely complex, we have showed that is necessary to benchmark QPUs on the basis of their variability for specific applications. Overall, we have shown that $\eta$ is a valuable metric which can be used to monitor the progress of NISQ QPUs as they evolve to produce higher quality approximate solutions to a wide array of combinatorial optimizations problems with the QAOA or indeed any combinatorial optimization algorithm whose result is a probability distribution of solutions to a decision problem. 

\section*{Acknowledgements}

 We thank Josiah Bjorgaard and Satish Gandhi of high-performance and quantum computing partner team at Amazon Web Services (AWS) for their collaboration, and acknowledge the use of Amazon Braket and AWS computational resources that made this work possible. We acknowledge the use of IBM Quantum services for this work. The views expressed are those of the authors, and do not reflect the official policy or position of IBM or the IBM Quantum team. We are grateful for the many useful conversations regarding this project with those at Agnostiq Inc. including Oktay Goktas, Elliot MacGowan, William Cunningham, Faiyaz Hasan, Haim Horowitz, Casey Jao, Anna Hughes, Jalani Kanem and Pooja Rao.


\bibliography{ms.bib}

\end{document}


\author{Jack S. Baker$^{1}$, \& Santosh Kumar Radha$^{1}$  \\\small{$^1$\textit{Agnostiq Inc., 325 Front St W, Toronto, ON M5V 2Y1}}}

\title{\textbf{Supplemental Material} \\ \medskip \Large{Wasserstein Solution Quality and the Quantum Approximate Optimization Algorithm: A Portfolio
Optimization Case Study}}
\maketitle

\medskip


This document contains the necessary supplemental material for the article entitled ``\textit{Wasserstein Solution Quality and the Quantum Approximate Optimization Algorithm: A Portfolio Optimization Case Study}". Requests for additional data and comments concerning the article are welcome via email contact\footnote[2]{Jack S. Baker: \textcolor{myurlcolor}{jack@agnostiq.ai}, Santosh Kumar Radha: \textcolor{myurlcolor}{santosh@agnostiq.ai}}.

\section{Number of API gates and QPU data summary}

Tabulations of the number of Application Programming Interface (API) gates (definition in the main text) used in Figure 10 in the main text compared with exactly the number of gates needed to implement the ans\"{a}tze at the ideal optimal angles $\boldsymbol{\beta}^*$ and $\boldsymbol{\gamma}^*$ (i.e, the optimal ans\"{a}tze). These are shown in Table \ref{tab:dickeandcomplete} and \ref{tab:ringandrandom}. A summary of the properties of each of QPU studied in the main text is included in Table \ref{tab:qpu_data}. \par

\section{Justifying the limited re-optimization on QPUs}

\begin{figure}
    \centering
    \includegraphics[width=\linewidth]{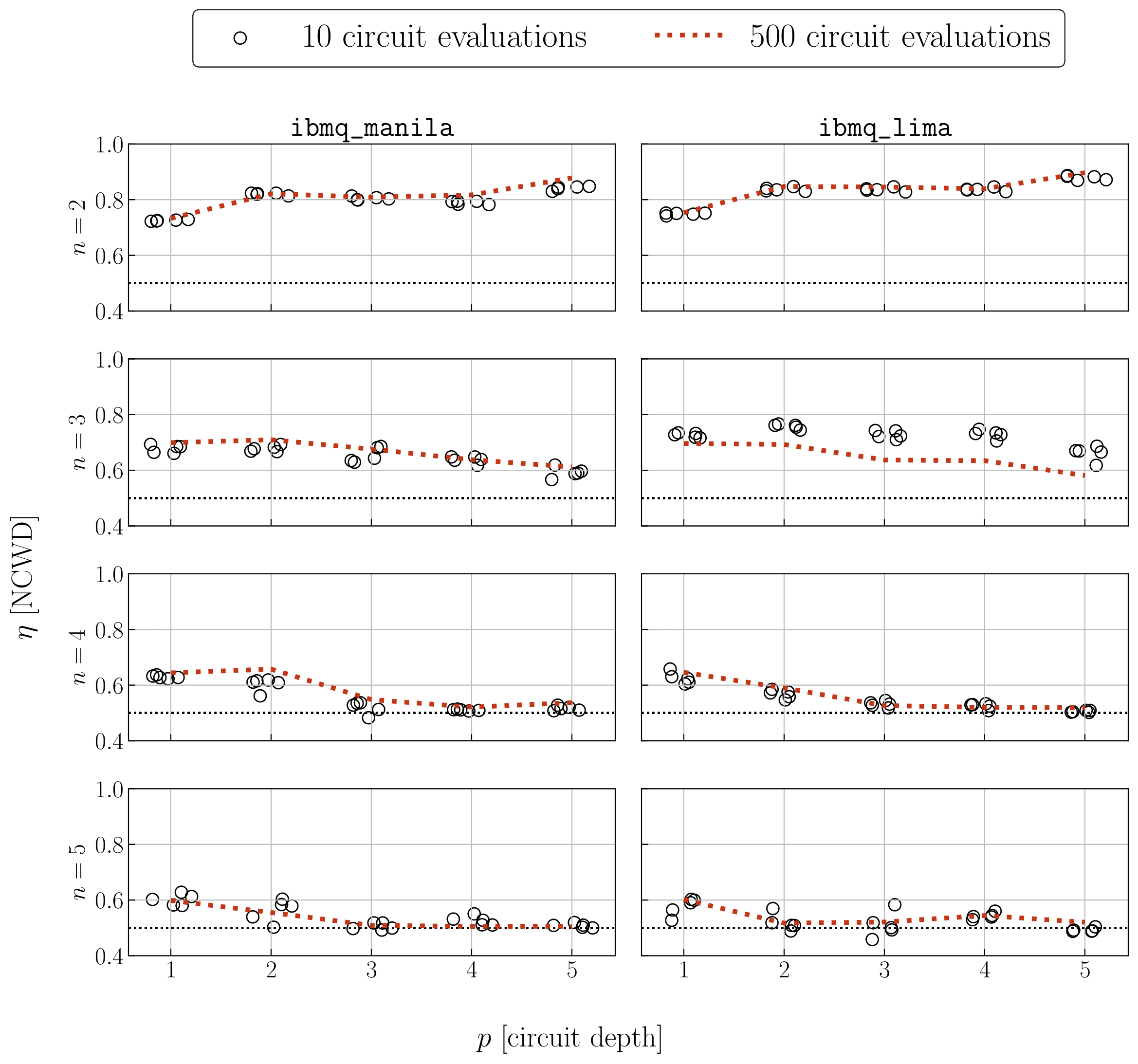}
    \caption{Limited re-optimization with 10 circuit evaluations (repeated 5 times) versus full re-optimization with 500 circuit evaluations on \texttt{ibmq\_manila} and \texttt{ibmq\_lima}. Both cases are warm started with the optimal angles from ideal execution. 10 circuit evaluation points are jittered slightly along the $x$-axis so all points are visible.}
    \label{fig:versus_500_shot}
\end{figure}

A justification for using a warm-started 10 circuit evaluations re-optimization is given in Figure \ref{fig:versus_500_shot}. That is, we compare the 5 repeats in the main text versus a full 500 circuit evaluation warm started optimization on the \texttt{ibmq\_manila} and \texttt{ibmq\_lima} QPUs. It can be seen, that on average, little-to-no improvement can be made using the larger number of circuit evaluations. One explanation for this is that the warm-start angles of the ideal simulator results correspond quite well with the same basin of attraction on the hardware; little correction is needed to approximate the minima on hardware. Interestingly, at $n = 3$ for \texttt{ibmq\_lima}, 500 circuit evaluations returns much worse results the 10 circuit evaluations. Indeed, as discussed in the main text, it could be that the general variability of the machine is greater than the possible benefits of a larger number of circuit evaluations. \par

\section{Additional black-box optimization data}

\begin{landscape}
\begin{figure}
    \centering
    \includegraphics[width=\linewidth]{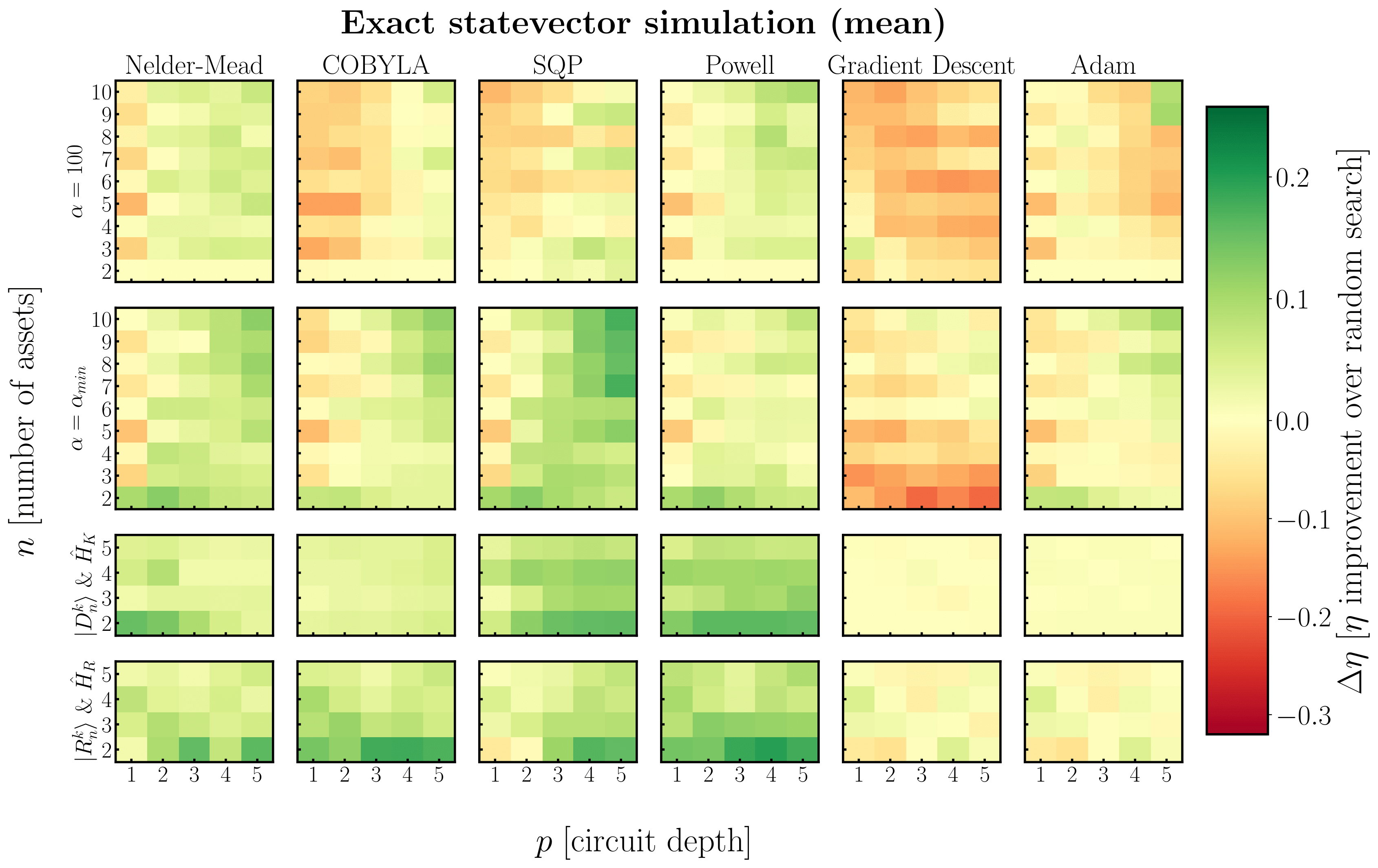}
    \caption{The mean value of $\eta$ achieved over all optimizer runs as compared to a random search over parameter space for exact statevector simulation. Columns (optimizer) and rows (constraint enforcement schemes) are described in the main text.}
    \label{fig:mean_sv}
\end{figure}
\end{landscape}

\begin{landscape}
\begin{figure}
    \centering
    \includegraphics[width=\linewidth]{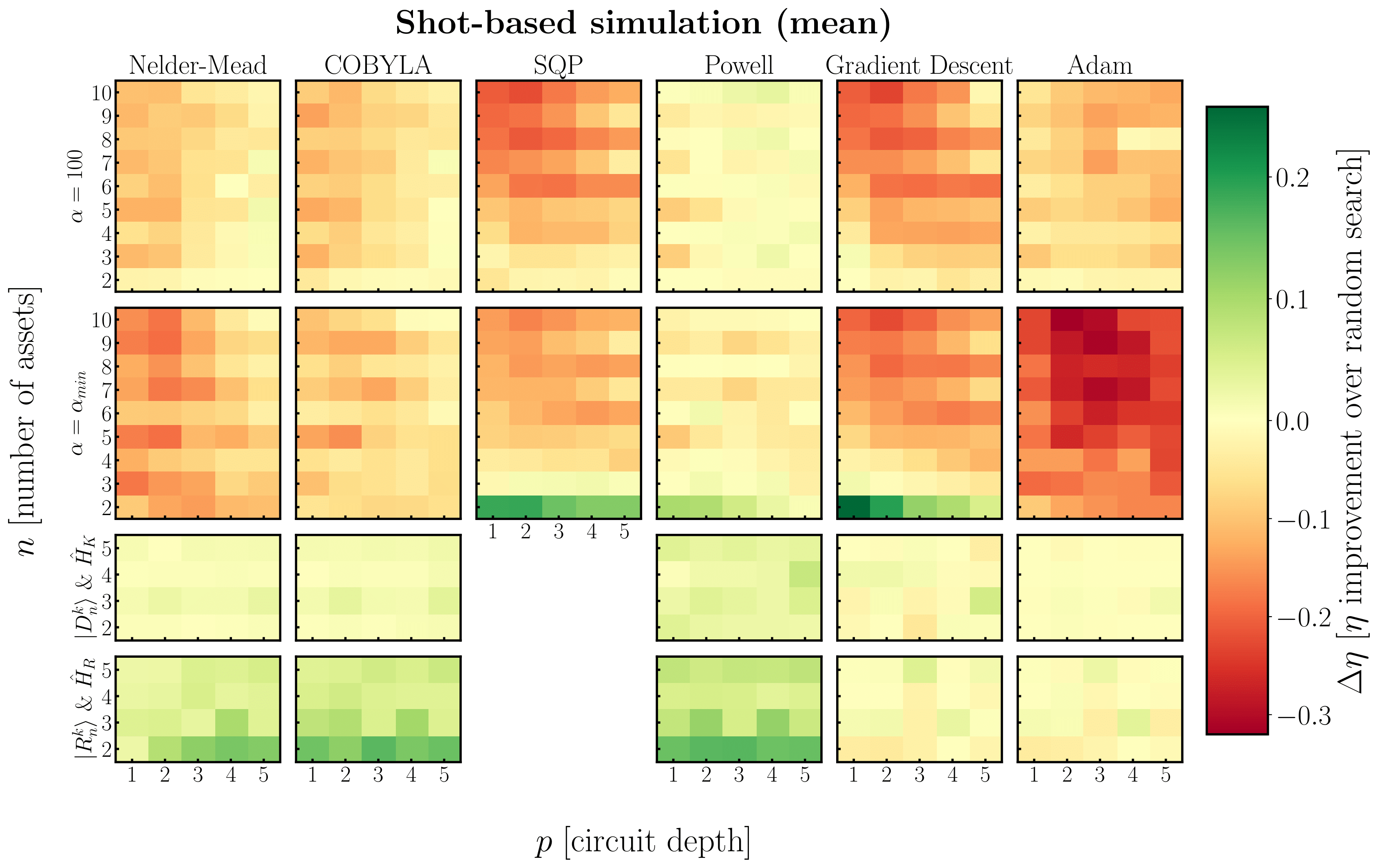}
    \caption{The mean value of $\eta$ achieved over all optimizer runs as compared to a random search over parameter space for shot-based simulation of 2048 shots. Columns (optimizer) and rows (constraint enforcement schemes) are described in the main text.}
    \label{fig:mean_shot}
\end{figure}
\end{landscape}

\begin{landscape}
\begin{figure}
    \centering
    \includegraphics[width=\linewidth]{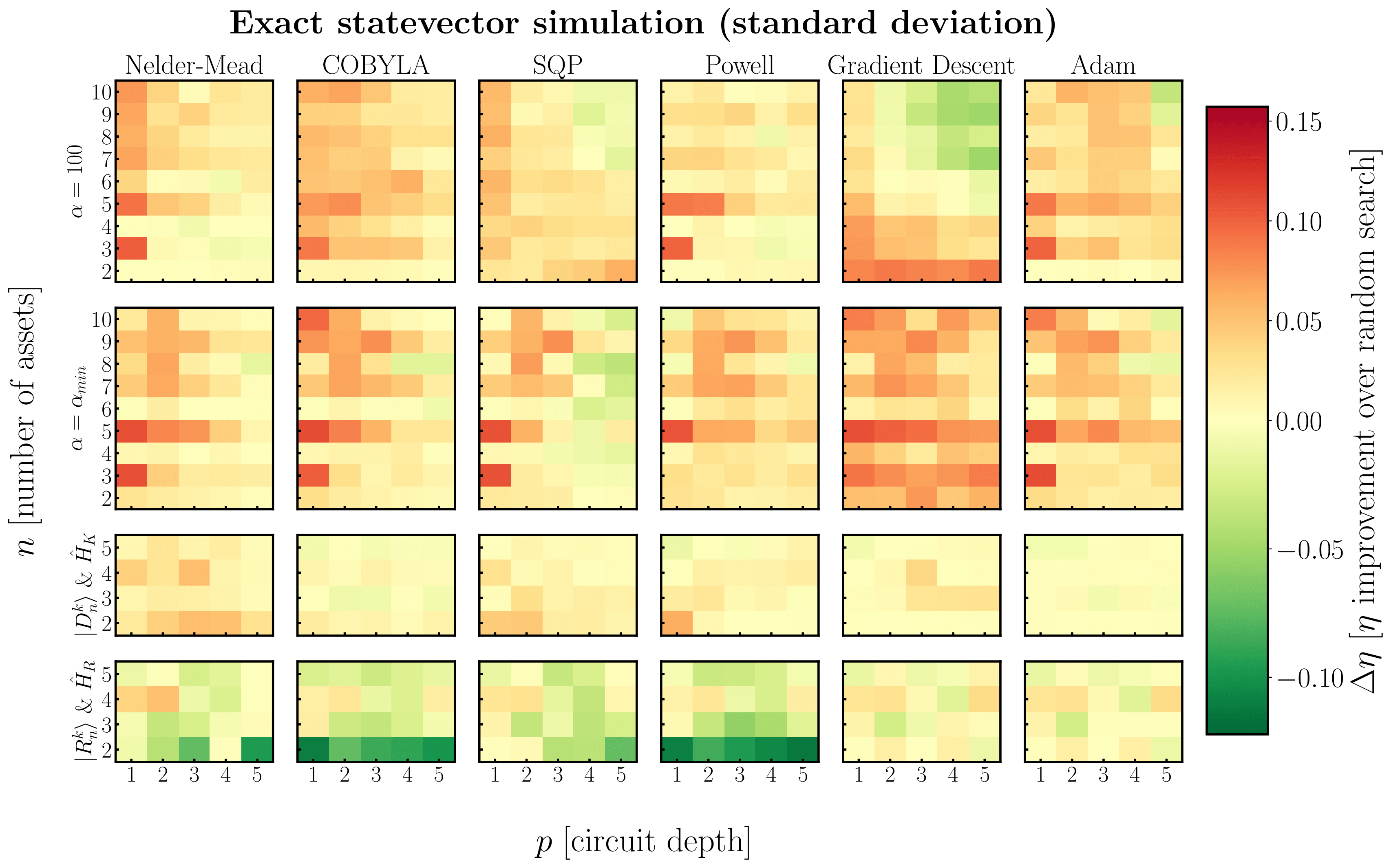}
    \caption{The standard deviation of $\eta$ from over all optimizer runs compared to a random search over parameter space for exact statevector simulations. Columns (optimizer) and rows (constraint enforcement schemes) are described in the main text.}
    \label{fig:std_sv}
\end{figure}
\end{landscape}

\begin{landscape}
\begin{figure}
    \centering
    \includegraphics[width=\linewidth]{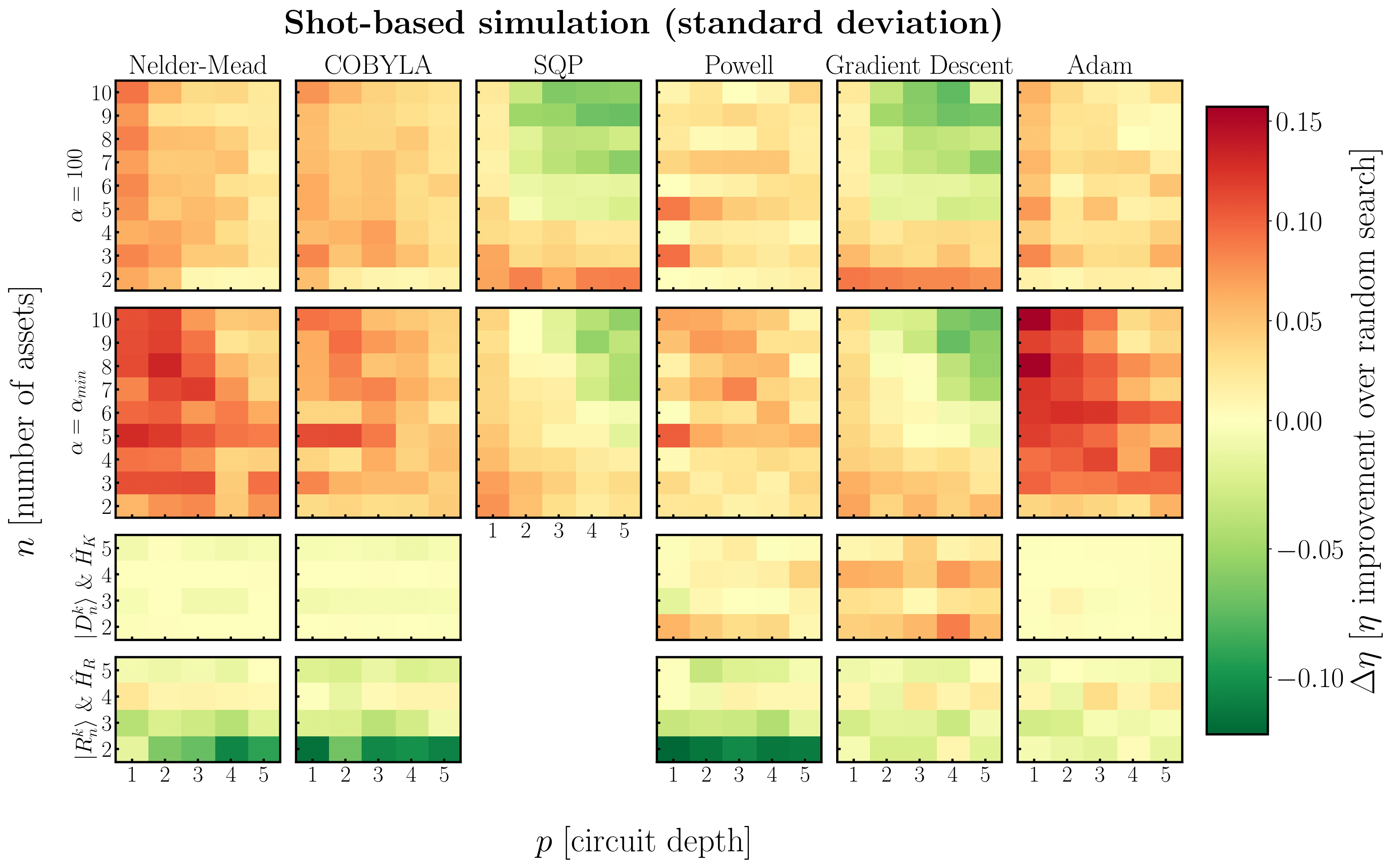}
    \caption{The standard deviation of $\eta$ from over all optimizer runs compared to a random search over parameter space for exact statevector simulations. Columns (optimizer) and rows (constraint enforcement schemes) are described in the main text.}
    \label{fig:std_shot}
\end{figure}
\end{landscape}

Additional data is provided for black-box optimizer benchmarks. Figure \ref{fig:mean_sv} and \ref{fig:mean_shot} show the average $\eta$ over all runs compared with random search for exact statevector and shot-based simulators, respectively. Figure \ref{fig:std_sv} and \ref{fig:std_shot} show the standard deviation of $\eta$ over all runs compared with random search for exact statevector and shot-based simulators, respectively. For Figure \ref{fig:std_sv} and \ref{fig:std_shot}, note that the color scale is inverted. I.e, a large standard deviation compared with random search is bad performance as far as reliability is concerned. \par

\bibliographystyle{ieeetr}
\bibliography{supplement.bib} 
\hrulefill

\medskip \medskip \medskip \medskip \medskip \medskip \medskip \medskip

\begin{table}[h]
\centering
\begin{tabular}{@{}cccccc@{}}
\toprule \toprule
        & \multicolumn{5}{c}{Average number of API gates/number of API gates at [$\boldsymbol{\gamma}^*$, $\boldsymbol{\beta}^*$ ]} \\ \midrule
        & $p = 1$              & $p = 2$               & $p = 3$              & $p = 4$              & $p = 5$              \\
$n = 2$ & 93/163               & 182/182               & 271/598              & 360/598              & 449/771              \\
$n = 3$ & 268/58               & 532/322               & 796/712              & 1060/682             & 1324/1114            \\
$n = 4$ & 530/530              & 1056/2232             & 1582/1834            & 2108/2108            & 2634/3054            \\
$n = 5$ & 879/1299             & 1754/3714             & 2629/3469            & 3504/1964            & 4379/5919            \\ \bottomrule
\end{tabular}
\caption{The average number of API gates over the bounds of $\boldsymbol{\beta}$ and the exact number of API gates at $\boldsymbol{\gamma}^*$ and $\boldsymbol{\beta}^*$ (the optimal angles of ideal execution) for $|D_n^B \rangle$ + $\hat{H}_K$.}
\label{tab:dickeandcomplete}
\end{table}

\begin{table}[h]
\centering
\begin{tabular}{@{}cccccc@{}}
\toprule
        & \multicolumn{5}{c}{Average number of API gates/number of API gates at [$\boldsymbol{\gamma}^*$, $\boldsymbol{\beta}^*$ ]} \\ \midrule
        & $p = 1$              & $p = 2$               & $p = 3$              & $p = 4$              & $p = 5$              \\
$n = 2$ & 93/163               & 182/182               & 271/598              & 360/598              & 449/771              \\
$n = 3$ & 268/58               & 532/322               & 796/712              & 1060/682             & 1324/1114            \\
$n = 4$ & 530/530              & 1056/2232             & 1582/1834            & 2108/2108            & 2634/3054            \\
$n = 5$ & 879/1299             & 1754/3714             & 2629/3469            & 3504/1964            & 4379/5919            \\ \bottomrule
\end{tabular}
\caption{The average number of API gates over the bounds of $\boldsymbol{\beta}$ and the exact number of API gates at $\boldsymbol{\gamma}^*$ and $\boldsymbol{\beta}^*$ (the optimal angles of ideal execution) for $|R_n^B \rangle$ + $\hat{H}_R$.}
\label{tab:ringandrandom}
\end{table}

\begin{landscape}
\begin{table}[]
\resizebox{1.285\textheight}{!}{%
\begin{tabular}{ccccccccc}
\toprule \toprule
Name             & Vendor  & Qubit type      & Scale & QV                & 1/2 qubit gate fidelities & $G_{WC}$ & Topology           & Native gates                \\ \hline
Aspen-10         & Rigetti & Superconducting & 32    & N/A               & 99.37/94.66               & 70.65    & Octagonal 2/3-fold & $RX$, $RZ$, $CZ$, $XY$    \\
11 qubit machine & IonQ    & Trapped Ion     & 11    & 64 {[}estimate{\cite{lubinski2021applicationoriented}]} & 99.76/97.19               & 952.38   & Complete graph     & $MS$, $GPI$, $GPI2$, $GZ$   \\
ibmq\_manila     & IBM     & Superconducting & 5     & 32                & 99.97/99.14               & 138.125  & 1D chain           & $CX$, $ID$, $RZ$, $SX$, $X$ \\
ibmq\_bogota     & IBM     & Superconducting & 5     & 32                & 99.98/97.58               & 248.99   & 1D chain           & $CX$, $ID$, $RZ$, $SX$, $X$ \\
ibmq\_quito      & IBM     & Superconducting & 5     & 16                & 99.97/99.05               & 378.26   & T shape            & $CX$, $ID$, $RZ$, $SX$, $X$ \\
ibmq\_belem      & IBM     & Superconducting & 5     & 16                & 99.97/99.01               & 148.54   & T shape            & $CX$, $ID$, $RZ$, $SX$, $X$ \\
ibmq\_lima       & IBM     & Superconducting & 5     & 8                 & 99.97/98.95               & 265.68   & T shape            & $CX$, $ID$, $RZ$, $SX$, $X$ \\ \hline
\end{tabular}%
}
\caption{The QPUs used in the study. Scale is the total number qubits and $G_{WC}$ is an estimate of the worst case number of implementable gates: $G_{WC} = \min{\{T_1, T_2\}}/ \max{\{T_{1Q}, T_{2Q}\}}$ where $T_1$ is the relaxation time, $T_2$ is the dephasing time, $T_{1Q}$ is the mean single qubit gate duration and $T_{2Q}$ is the mean 2 qubit gate duration.}
\label{tab:qpu_data}
\end{table}
\end{landscape}

